# Full Poincaré-invariant equation of motion for an extended charged particle in an electromagnetic field

**Helmut Stöckel**

Haus 5, 18279 Alt-Krassow (Germany)[1]

## Abstract

For the rigid, nonrotating motion of an extended charge in an arbitrary electromagnetic field, an equation of motion is derived by Lorentz-invariantly calculating the 4-Lorentz force $\bar{F} = \bar{F}_e + \bar{F}_s =$ external 4-force + 4-self-force, acting upon the charge. For small 4-acceleration $\bar{a}$, the 4-self-force becomes $\bar{F}_s \approx -m_1 \bar{a}$, where $m_1 c^2$ equals the electrostatic self-energy. The notorious factor 4/3 does not occur.

The equation of motion is invariant under any full Poincaré transformation. It is a differential-integral equation for the particle's 4-centre $\bar{z}$ in Minkowski space as a function of proper time $\tau$. The twofold proper time integrated form of the equation of motion is suited to derive the solution $\bar{z}(\tau)$ iteratively if the initial conditions $\bar{z}(0) = \bar{z}_0$, $\dot{\bar{z}}(0) = \dot{\bar{z}}_0$ and the external electromagnetic field are given and the iteration procedere converges. Diverse approximations for the self-force and with that also for the equation of motion are derived.

The (approximated) equation of motion is solved for two examples: (1) Rectilinear hyperbolic motion in a homogeneous electric field. (2) Spiral (nonrelativistic) motion in a homogeneous magnetic field. In both cases amazingly simple solutions result.

The advantages of the equation against other ones, for example the Lorentz-Dirac equation, are discussed.

---

[1] helmut.stoeckel@web.de

# 1. Introduction

> *A physicist needs that his equations should be mathematically sound and that in working with his equations he should not neglect quantities unless they are small.*
> P.A.M. Dirac [13]

*How does a charged particle move in an electromagnetic field under consideration of the reaction force?*

Enormous efforts to answer this question have been made for more than a century. The history of these efforts may be looked up in the following selected literature: [1,3,4,6,8,12,13, 14,15,20,21,25,27,28,30,35,40,44,47,49,50,54,59,62,63,66]. In these works the drawbacks of the various equations of motion and their solutions are described in detail (see also Appendix F). A further trial to answer this question shall be made in the following.

As the assumption of a point charge leads to mathematically and physically unsatisfying consequences [24,43,44,48,63], we suppose an extended charge with spherically symmetric charge density. The particle shall execute a rigid nonrotating motion [9,18,26,41,64,65]. At sufficiently small particle radius (4.9) the spherically symmetric charge density distribution in the rest frame is conserved under the transition to the *local* rest frame.

To derive the Lorentz-invariant equation of motion (8.2), we first explain the conception of a *rigid nonrotating motion*. We prefer the compact notation of 4-vectors and tensors against the more frequently used coordinate notation, as the former is more lucid. The compact vectors and tensors (dyadics) are Lorentz-invariant, their coordinate representations are only forminvariant (covariant).

The unknown density function $\rho(r)$ (4.1) of the particle is chosen so that the square $\tilde{\rho}^2$ of its Fourier transform (4.6) is as simple as possible. The density $\rho$ is used as weight factor to calculate the mean values of the 4-current density $\overline{j}$, of the retarded 4-self-potential $\overline{A}_s$, and of the electromagnetic 4-self-field strength $\overline{\overline{B}}_s$ (7.1a) (4-dyadic = compact representation of the electromagnetic field strength tensor). The space part $\vec{B}_s$ of $\overline{\overline{B}}_s$ (7.1c) is linearly connected with the magnetic self-field strength $\vec{B}_s$. We prefer the letter $B$ against the frequently used letter $F$ to designate the electromagnetic field strength, in analogy to $\overline{j} = j^0 \overline{e}_0 + \vec{j}$ and $\overline{A} = A^0 \overline{e}_0 + \vec{A}$ [cf. (6.7)], and as the letter $F$ is used to designate forces.

Additionally to the mean electromagnetic self-field strength $<\overline{\overline{B}}_s>$ (7.6), the external electromagnetic field strength $\overline{\overline{B}}_e$ acts on the particle. The sum of both, $<\overline{\overline{B}}> := \overline{\overline{B}}_e + <\overline{\overline{B}}_s>$, causes the 4-Lorentz force $\overline{F}$ in the equation of motion (8.1) or (8.2). Twofold proper time integration of (8.2) yields the equivalent differential-integral equation (8.5). This is suited to derive the solution $\overline{z}(\tau)$ iteratively if the initial conditions (8.4) and the external electromagnetic field strength $\overline{\overline{B}}_e(\overline{x})$, with $\overline{x} = t\overline{e}_0 + x^1\overline{e}_1 + x^2\overline{e}_2 + x^3\overline{e}_3$ (2.6), are given. The 4-position vector $\overline{z}(\tau)$ describes the worldline of the particle centre with the proper time as a parameter. Various approximations for the self-force and with it for the equation of motion are derived.

The approximated equation of motion is solved for two examples: (1) The rectilinear (hyperbolic) motion of a charge $q$ in a homogeneous electric field $E_e = const$. (2) The spiral (nonrelativistic) motion of $q$ in a homogeneous magnetic field $B_e = const$. In both cases, the solutions $\overline{z}(\tau)$ satisfy the generalized Newton law $m\,a = F_e$.

The generalization consists in that the *effective mass* $m$ becomes a (generally nonanalytic) function of $|a|$, where $a := \ddot{z}(\tau)$ is the proper acceleration. In case of $B_e = const$, it is advantageous, to assume $m$, $a$ and $F_e$ as complex quantities. The notorious factor 4/3 does not occur.

For simplicity, we set $c = \varepsilon_0 = \mu_0 = 1$. For the rest, we use the rational international quantity system.

The equation of motion (8.2) is invariant under any full Poincaré transformation (= full inhomogeneous Lorentz transformation) of coordinates *and* base vectors [see Sect. 19]. By each passive transformation [22,56], the *physical processes shall not be transformed*, but only their description as to another orthonormal base $\{\bar{e}_i\}$. *A Lorentz transformation is always a passive transformation* [22].

A time inversion $t \to -t$, $\bar{e}_0 \to -\bar{e}_0$ is a special transformation of the full Poincaré group. It lets each 4-point $\bar{x} = x^i \bar{e}_i$, $i = 0, ..., 3$ (sum convention), invariant. For example, by a passive time inversion an inward spiraling charge $q$ is not transformed into an outward spiraling one. Instead of this, a backward running clock describes the inward spiraling process.

For the spiraling particle, power balances are discussed. General discussions of 4-momentum balances and of the energy-momentum tensor are omitted. For, they are in detail discussed in the text books (see, e.g., [20,22,28,34,42,54,56,57,60,65]).

In the Appendix F the Lorentz-Dirac equation is derived as an approximation of the equation of motion (8.2). Unfortunately, the Laurent expansion (F4) of the 4-self-force $\bar{F}_s$ and the truncation of this in general divergent, asymptotic expansion are inadmissible. These unsound mathematical operations obviously cause the unsound results.

At the end, advantages of the equation of motion (8.2), compared with other ones, are discussed. In the Appendices A to F various proofs and supplements are given.

## 2. Rigid nonrotating motion of a charged particle

Laxly spoken, we'll describe the motion of a rigid, nonrotating charged particle in an electromagnetic field. How is this Lorentz-invariantly possible under consideration of the particle's self-field? For simplicity, we assume that the particle's charge centre coincides with its mass centre.

The 4-position of the centre may be represented by

$$\bar{z} = z^i \bar{e}_i. \tag{2.1}$$

(Sum convention: Over double Latin indices $i, j, ...$ is to be summed from 0 to 3, over Greek indices $\alpha, \beta, ...$ from 1 to 3; like, e.g., in [34].) The $z^i$ are the contravariant coordinates of $\bar{z}$ with respect to the (covariant) 4-base $\{\bar{e}_i\}$ of the lab frame. The orthonormal unit vectors $\bar{e}_i$ point respectively into the t, x, y, z directions. They define the Lorentz (lab) metric

$$(\bar{e}_i \cdot \bar{e}_j) =: (\eta_{ij}) = (\eta^{ij}) := diag(1, -1, -1, -1). \tag{2.2a}$$

(The symbols $:=$ and $=:$ mean: *equals by definition*. The colon stands on the side of the defined quantity.)

The contravariant (dual) base $\{\bar{e}^i\}$ is defined by the orthonormality relations

$$(\bar{e}_i \cdot \bar{e}^j) =: (\eta_i{}^j) := (\delta_i{}^j) = diag(1, 1, 1, 1). \tag{2.2b}$$

Hence, $\bar{e}^0 = \bar{e}_0$, $\bar{e}^\alpha = -\bar{e}_\alpha$.

The 4-vector $\bar{z}$ depends on its proper time $\tau$. The 4-velocity

$$\frac{d\bar{z}}{d\tau} =: \dot{\bar{z}}(\tau) =: \bar{u}, \quad \text{with} \quad \bar{u}^2 := \bar{u}\cdot\bar{u} = (u^0)^2 - u_\alpha u^\alpha = 1, \tag{2.3}$$

is a time-like unit vector.

Additionally to the lab base $\{\bar{e}_i\}$, we introduce the primed base

$$\{\bar{e}'_i\} \quad \text{with} \quad \bar{e}'_i \cdot \bar{e}'_j = \eta_{ij}, \quad \bar{e}'_0 := \bar{u}(\tau) \tag{2.4}$$

of the local (instantaneous) rest frame (= inertial frame) at a 4-point $\bar{z}(\tau)$ of a worldline. The 4-vector $\bar{z}$ may also be represented with respect to this base:

$$\bar{z} = z'^i \bar{e}'_i. \tag{2.5}$$

The same holds for an arbitrary 4-point of the particle:

$$\bar{x} = x^i \bar{e}_i = x'^i \bar{e}'_i = x_i \bar{e}^i = x'_i \bar{e}'^i. \tag{2.6}$$

We speak from a *rigid nonrotating motion* [9,18,26,41,64,65] if $\bar{x}(\tau)$ is connected with $\bar{z}(\tau)$ as follows:

$$\bar{x}(\tau) := \bar{z}(\tau) + r^\alpha \bar{e}'_\alpha(\tau) = \bar{z}(\tau) + \bar{r}(\tau), \quad r^\alpha = const, \tag{2.7a}$$

$$\frac{d\bar{x}}{d\tau} =: \dot{\bar{x}} := \lambda \dot{\bar{z}}, \quad \lambda \text{ scalar}. \tag{2.7b}$$

That is, the particle fixed position vector

$$\bar{r} := \bar{x} - \bar{z} = r^\alpha \bar{e}'_\alpha(\tau) \tag{2.8}$$

lies in the 3-plane

$$\sigma := \{\bar{r} \mid \bar{r} \cdot \dot{\bar{z}}(\tau) = 0\} \tag{2.9}$$

**Fig. 1** through $\bar{z}$ orthogonal to $\dot{\bar{z}}$ (Fig. 1).

## 3. Functional determinant

Corresponding to (2.7a), the 4-point $\bar{x}$ is also determined by the coordinates $\tau =: r^0, r^1, r^2, r^3$. These are connected with the lab coordinates $x^i$ by

$$x^i = z^i(\tau) + r^\alpha e'^i_\alpha(\tau). \tag{3.1}$$

The functional determinant of this coordinate transformation reads

$$D := \det\left(\frac{\partial x^i}{\partial r^j}\right) = \begin{vmatrix} \dot{z}^0 + r^\alpha \dot{e}'^0_\alpha & e'^0_1 & e'^0_2 & e'^0_3 \\ \dot{z}^1 + r^\alpha \dot{e}'^1_\alpha & e'^1_1 & e'^1_2 & e'^1_3 \\ \dot{z}^2 + r^\alpha \dot{e}'^2_\alpha & e'^2_1 & e'^2_2 & e'^2_3 \\ \dot{z}^3 + r^\alpha \dot{e}'^3_\alpha & e'^3_1 & e'^3_2 & e'^3_3 \end{vmatrix}. \tag{3.2}$$

For a local rest base holds $e'^0_1 = e'^0_2 = e'^0_3 = 0$ [see (2.4) and Fig. 1], so that this determinant specializes to

$$D = 1 + r^\alpha \dot{e}'^0_\alpha = 1 + r^\alpha \dot{\bar{e}}'_\alpha \cdot \dot{\bar{z}} \stackrel{(2.8)}{=} 1 + \dot{\bar{r}} \cdot \dot{\bar{z}} \equiv 1 - \bar{r} \cdot \ddot{\bar{z}}. \tag{3.3}$$

(The equation number (2.8) above the equality sign refers to the cited equation.) The last identity follows from

$$\bar{r} \cdot \dot{\bar{z}} \stackrel{(2.9)}{=} 0 \tag{3.4}$$

by differentiation with respect to $\tau$. As the result (3.3) is Lorentz-invariant, it also holds with respect to the lab base.

*The factor of proportionality $\lambda$ in (2.7b) equals $D$ (3.3).* To prove this assertion, we equate (2.7b) with the proper time derivative of (2.7a):

$$\dot{\bar{x}} = \lambda \dot{\bar{z}} = \dot{\bar{z}} + \dot{\bar{r}}.$$

Scalar multiplication with $\dot{\bar{z}}$ yields

$$\lambda = 1 + \dot{\bar{r}} \cdot \dot{\bar{z}} = D, \quad \text{q.e.d.} \tag{3.5}$$

On condition that

$$<r> a \ll 1, \quad a := |\bar{a}| := |\ddot{\bar{z}}|, \tag{3.6}$$

(cf. (14.13)), the functional determinant can very well be approximated by

$$D \approx 1. \tag{3.7}$$

Here $<r>$ means the particle radius (4.9). Then the 4-volume elements $d^4\bar{x}$ and $d^4\bar{r}$ are approximately equal:

$$d^4\bar{x} := dx^0 dx^1 dx^2 dx^3 = D d^4\bar{r} := D d\tau\, dr^1 dr^2 dr^3$$
$$\approx d\tau dr^1 dr^2 dr^3 =: d\tau d\sigma. \tag{3.8}$$

The 3-volume element $d\sigma = dr^1 dr^2 dr^3$ is identical with the surface element of the 3-plane $\sigma$ (2.9).

## 4. Density

For simplicity, let us assume that the *density*

$$\rho := \text{charge density} / \text{charge} = \rho_{el}/q \tag{4.1}$$

in the particle's rest frame is spherically symmetric:

$$\rho = \rho(r) \quad \text{with} \quad r := |\bar{r}| := \sqrt{-\bar{r}^2} = \sqrt{(r^1)^2 + (r^2)^2 + (r^3)^2}. \tag{4.2}$$

On condition (3.6), this also holds in each local rest frame.

In the following it proves favorable to transform the density $\rho(r)$ into its Lorentz-invariant 3-Fourier transform (structure function)

$$\tilde{\rho}(\kappa) := \int_\sigma d\sigma \exp(\mp i \bar{k} \cdot \bar{r})\, \rho(r) = \frac{4\pi}{\kappa} \int_0^\infty dr\, r \sin \kappa r\, \rho(r) = \tilde{\rho}(-\kappa), \tag{4.3a}$$

with

$$\kappa := \sqrt{(\bar{k} \cdot \dot{\bar{z}})^2 - \bar{k}^2}, \quad \tilde{\rho}(0) = \int_\sigma d\sigma\, \rho(r) = 1. \tag{4.3b}$$

In the (primed) local rest frame holds $\dot{\bar{z}} = \vec{e}_0'$ (2.4), with which we get

$$\kappa = \sqrt{(k'^1)^2 + (k'^2)^2 + (k'^3)^2} \overset{(6.7)}{=:} |\vec{k}'|. \tag{4.3c}$$

The reversal of (4.3a) is

$$\rho(r) = \frac{1}{i4\pi^2 r} \int_{-\infty}^\infty d\kappa\, \kappa \exp(i\kappa r) \tilde{\rho}(\kappa) = \frac{1}{4\pi^2 r} \int_{-\infty}^\infty d\kappa\, \kappa\, \tilde{\rho}(\kappa) \sin \kappa r = \rho(-r). \tag{4.4}$$

For a point charge holds

$$\rho(r) = \frac{\delta(r)}{2\pi r^2}, \quad \tilde{\rho} = 1. \tag{4.5}$$

Later we will specialize on the simple Fourier transform

$$\tilde{\rho}_1(\kappa) := (1 + r_1^2 \kappa^2)^{-1/2}. \tag{4.6}$$

The reason for this choice is that $\tilde{\rho}_1^2(\kappa)$ has only poles of first order at $\kappa = \pm i/r_1$ as singularities in the complex $\kappa$-plane. The parameter $r_1$ has the meaning of a quasi-radius. The comparison with [23,31] suggests identifying $r_1$ with half the Compton wavelength:

$$r_1 := \frac{1}{2}\lambda_C := \frac{\hbar}{2m_0 c}. \tag{4.7}$$

The insertion of (4.6) into (4.4) yields the *density function*

$$\rho_1(r) = \frac{1}{2\pi^2 r \, r_1^2} K_1\!\left(\frac{r}{r_1}\right) \tag{4.8}$$

with

$$K_1(x) := \int_0^\infty dz \, \cosh z \, \exp(-x \cosh z)$$

**Fig. 2**    (cf. [2,19,29,60] and Fig. 2).

As *particle radius* we define the mean value of $r$ in the particle's rest frame:

$$\langle r \rangle := \int_\sigma d\sigma \, \rho(r) r = 4\pi \int_0^\infty dr \, r^3 \rho(r). \tag{4.9}$$

Expressing $\rho(r)$ by $\tilde{\rho}(\kappa)$ (4.4) and integrating over $r$ gives

$$\langle r \rangle = \frac{2}{\pi} \int_{-\infty}^\infty d\kappa \frac{\tilde{\rho}}{(\kappa + i0)^2} := \lim_{\varepsilon \to 0} \frac{2}{\pi} \int_{-\infty}^\infty d\kappa \frac{\tilde{\rho}}{(\kappa + i\varepsilon)^2}. \tag{4.10}$$

For the special form factor (4.6) one gets

$$\langle r \rangle = \frac{4 r_1}{\pi}. \tag{4.11}$$

## 5. Mean four-current density

The *4-current density* of the particle with the charge $q$ is

$$\overline{j} = q \rho(r) \dot{\overline{x}}(\tau) \overset{(3.7)}{\approx} q \rho \dot{\overline{z}}(\tau) \quad \Rightarrow \quad \rho = \frac{1}{q}\sqrt{\overline{j}^2}. \tag{5.1}$$

The 4-Fourier transform becomes

$$\overline{j}_{\overline{k}} = \int d^4\overline{x} \, \exp(i\overline{k} \cdot \overline{x}) \, \overline{j}. \tag{5.2}$$

The 4-volume integral is to be taken over that domain of spacetime in which $\overline{j} \neq \overline{0}$. Because $\rho(r)$ is $\delta$-distribution like, only volume elements $d^4\overline{x}$ within a thin tube of radius $\langle r \rangle$ about the worldline $\overline{z}(\tau)$ essentially contribute to the integral. Therefore, for not too large acceleration $a$, the inequality (3.6) is fulfilled, and $d^4\overline{x} \approx d\tau \, d\sigma$ (3.8) is a very good approximation:

$$\overline{j}_{\overline{k}} \approx q \int_{-\infty}^\infty d\tau \, \dot{\overline{z}} \exp(i\overline{k} \cdot \overline{z}) \int_\sigma d\sigma \, \exp[i\overline{k} \cdot (\overline{x} - \overline{z})] \rho(|\overline{x} - \overline{z}|)$$

$$\overset{(4.3a)}{=} q \int d\overline{z} \, \exp(i\overline{k} \cdot \overline{z}) \tilde{\rho}(\kappa). \tag{5.3}$$

In case (4.5) of a point charge, $\overline{j}_{\overline{k}}$ simplifies to the integral

$$\overline{j}_{\overline{k}} = q \int_{\overline{z}(-\infty)}^{\overline{z}(\infty)} d\overline{z} \, \exp(i\overline{k} \cdot \overline{z}) \quad \text{for} \quad \tilde{\rho} = 1 \tag{5.4}$$

along the worldline $\overline{z}(\tau)$ [22].

In analogy to the radius $<r>$ (4.9), we define the *mean 4-current density* $<\bar{j}>$ as the mean value of the 4-current density $\bar{j}(\bar{x})$, averaged over the 3-plane $\sigma$ (2.9) with the density $\rho$ as a weight factor:

$$<\bar{j}>(\bar{z}) := \int_\sigma d\sigma\, \rho\, \bar{j} = \int_\sigma d\sigma\, \rho(|\vec{x}-\vec{z}|)\, C^4 \int d^4\bar{k}\, \exp(-i\bar{k}\cdot\bar{x})\, \bar{j}_{\bar{k}}$$

$$= C^4 \int d^4\bar{k}\, \exp(-i\bar{k}\cdot\bar{z})\, \bar{j}_{\bar{k}} \int_\sigma d\sigma\, \rho(|\vec{x}-\vec{z}|)\, \exp[-i\bar{k}\cdot(\bar{x}-\bar{z})]$$

$$\stackrel{(4.3a)}{=} C^4 \int d^4\bar{k}\, \exp(-i\bar{k}\cdot\bar{z})\, \bar{j}_{\bar{k}}\, \tilde{\rho}(\kappa) = C^4 \int d^4\bar{k}\, \exp(-i\bar{k}\cdot\bar{z}) <\bar{j}>_{\bar{k}}, \qquad (5.5a)$$

with

$$C := \frac{1}{2\pi}, \qquad <\bar{j}>_{\bar{k}} := \bar{j}_{\bar{k}}\, \tilde{\rho}(\kappa) \qquad (5.5b,c)$$

and $\kappa$ corresponding to (4.3b). This simple result signifies: The 4-Fourier transform of $<\bar{j}>$ equals the 4-Fourier transform of $\bar{j}$, multiplied by the 3-Fourier transform of the density $\rho$. The rule (5.5c) also holds for any other quantity (scalar, 4-vector, 4-tensor) instead of $\bar{j}$.

## 6. Mean retarded four-self-potential

The 4-current density

$$\bar{j}(\bar{x}) = C^4 \int d^4\bar{k}\, \exp(-i\bar{k}\cdot\bar{x})\, \bar{j}_{\bar{k}} \qquad (6.1)$$

of the moving charge generates the Lorenz-gauged [28] *4-self-potential* [22,34]

$$\bar{A}_s(\bar{x}) = \partial_{\bar{x}}^{-2}\bar{j} = C^4 \int d^4\bar{k}\, \exp(-i\bar{k}\cdot\bar{x})\, \frac{\bar{j}_{\bar{k}}}{-\bar{k}^2}, \qquad \partial_{\bar{x}}\cdot\bar{A}_s = \partial_{\bar{x}}\cdot\bar{j} = 0. \qquad (6.2)$$

Here $\partial_{\bar{x}}^{-2} = \square^{-1}$ means the inverse wave operator and

$$\partial_{\bar{x}} := \bar{e}_i \frac{\partial}{\partial x_i} =: \bar{e}_i \partial^i, \quad \text{with} \quad x_i := \eta_{ij} x^j, \qquad (6.3)$$

is the 4-nabla operator.

The *mean retarded 4-self-potential* follows analogously to (5.5a), substituting there $\bar{j}_{\bar{k}}$ by $\bar{A}_{s\bar{k}} = -\bar{j}_{\bar{k}}/\bar{k}^2$:

$$<\bar{A}_s>(\bar{z}) = C^4 \int d^4\bar{k}\, \exp(-i\bar{k}\cdot\bar{z})\frac{\bar{j}_{\bar{k}}}{-\bar{k}^2}\tilde{\rho}(\kappa) = \partial_{\bar{z}}^{-2} <\bar{j}>(\bar{z}). \qquad (6.4)$$

Inserting (5.3) we get

$$<\bar{A}_s>(\bar{z}) = -qC^4 \int_{-\infty}^{\infty} d\tau'\, \dot{\bar{z}}' \int d^4\bar{k}\, \exp[i\bar{k}\cdot(\bar{z}'-\bar{z})]\, \frac{\tilde{\rho}\tilde{\rho}'}{\bar{k}^2}, \qquad (6.5)$$

with

$$\bar{z}' := \bar{z}(\tau'), \quad \dot{\bar{z}}' := d\bar{z}'/d\tau',$$

$$\tilde{\rho} := \tilde{\rho}(\kappa) := \tilde{\rho}\left(\sqrt{(\bar{k}\cdot\dot{\bar{z}})^2 - \bar{k}^2}\right), \qquad \tilde{\rho}' := \tilde{\rho}(\kappa') := \tilde{\rho}\left(\sqrt{(\bar{k}\cdot\dot{\bar{z}}')^2 - \bar{k}^2}\right). \qquad (6.6)$$

We now decompose the 4-wave vector $\bar{k}$, the particle's 4-position $\bar{z}(\tau)$ and $\bar{z}' := \bar{z}(\tau')$ into (base-dependent) *time- and space-components* with respect to the lab base $\{\bar{e}_i\}$:

$$\bar{k} = k^i \bar{e}_i = \omega \bar{e}_0 + \vec{k}, \quad \bar{z} = z^0 \bar{e}_0 + \vec{z}, \quad \bar{z}' = z'^0 \bar{e}_0 + \vec{z}'. \qquad (6.7)$$

We distinguish the *dot product between 4-vectors* of Minkowski space from that *between 3-vectors* of the spatial Euklidean subspace $\perp \vec{e}_0$ by $\cdot$ and $\bullet$, respectively:

$$\bar{k} \cdot \bar{z} := k^0 z^0 - \vec{k} \bullet \vec{z} = k^0 z^0 - k^\alpha z^\alpha. \tag{6.8}$$

In the manner of speaking we distinguish between the (scalar) time coordinate $k^0 = \omega$ of $\bar{k}$ and the (vectorial) time component $k^0 \bar{e}_0$. The *space component* $\vec{k}$ of $\bar{k}$ has the *space coordinates* $k^\alpha$. If $k^0 = 0$, then $\bar{k} = \vec{k}$; cf. (10.1). With that, (6.5) reads

$$<\bar{A}_s> \approx -qC^4 \int_{-\infty}^{\infty} d\tau' \, \dot{\bar{z}}' \int_{C_r} d\omega \int d^3\vec{k} \, \exp\{i[\omega(z'^0 - z^0) - \vec{k}\bullet(\vec{z}' - \vec{z})]\} \, \frac{\tilde{\rho}\tilde{\rho}'}{\omega^2 - \vec{k}^2}. \tag{6.9}$$

**Fig. 3** The contour $C_r$ of integration in the complex $\omega$-plane is to be chosen like visualized in Fig. 3 so that $<\bar{A}_s>$ becomes the *retarded 4-potential* belonging to $<\bar{j}>$ [22,38,54]. The poles of the integrand at

$$\omega = \pm |\vec{k}| \tag{6.10}$$

and eventual singularities of $\tilde{\rho}'$ (cf. (9.4a,b)) are by-passed above. Consequently, the $\omega$-integral vanishes for times $z'^0 > z^0$. That is, $<\bar{A}_s>$ is causal, and the upper border of the proper time integral may be replaced by $\tau$:

$$\int_{-\infty}^{\infty} d\tau' \, \dot{\bar{z}}' \ldots = \int_{-\infty}^{\tau} d\tau' \, \dot{\bar{z}}' \ldots = \int_{-\infty}^{z^0} dz'^0 \, \frac{d\bar{z}'}{dz'^0} \ldots = \int_{\bar{z}'(-\infty)}^{\bar{z}} d\bar{z}' \ldots . \tag{6.11}$$

## 7. Mean electromagnetic self-field strength

From the 4-self-potential (6.2), the *electromagnetic 4-self-field strength* follows as the antisymmetric tensor (dyadic) [38]

$$\bar{\bar{B}}_s := -\bar{\partial}_{\bar{x}} \bar{A}_s + \bar{A}_s \bar{\partial}_{\bar{x}} = -iC^4 \int_{C_r} d^4\bar{k} \, \exp(-i\bar{k}\cdot\bar{x}) \, \frac{\bar{k}\,\bar{j}_{\bar{k}} - \bar{j}_{\bar{k}}\,\bar{k}}{\bar{k}^2}. \tag{7.1a}$$

Here $\bar{\partial}_{\bar{x}}$ means the *4-nabla differential operator* acting upon the left:

$$\bar{\partial}_{\bar{x}} := \frac{\bar{\partial}}{\partial x_i} \bar{e}_i =: \bar{\partial}^i \bar{e}_i. \tag{7.1b}$$

The electromagnetic 4-self-field strength $\bar{\bar{B}}_s$ linearly depends on the electric and magnetic self-field strengths $\vec{E}_s$ and $\vec{B}_s$ [22]:

$$\bar{\bar{B}}_s = \bar{e}_0 \vec{E}_s - \vec{E}_s \bar{e}_0 + \vec{\bar{B}}_s, \qquad \vec{\bar{B}}_s := -\vec{\mathbb{1}} \times \vec{B}_s = -\vec{B}_s \times \vec{\mathbb{1}}, \qquad \vec{\mathbb{1}} := \bar{e}_\alpha \bar{e}_\alpha. \tag{7.1c}$$

The contravariant coordinates $B_s^{ij}$ of the 4-dyadic $\bar{\bar{B}}_s$ with respect to the covariant base vectors $\bar{e}_i$ are defined by the expansion

$$\bar{\bar{B}}_s = B_s^{ij} \bar{e}_i \bar{e}_j = (-\partial^i A_s^j + A_s^i \bar{\partial}^j) \bar{e}_i \bar{e}_j. \tag{7.2}$$

Left-multiplying by $\bar{e}_0$ yields the *electric self-field strength*:

$$\bar{e}_0 \cdot \bar{\bar{B}}_s = \left(-\frac{\partial}{\partial x_0} A_s^j + \frac{\partial}{\partial x_j} A_s^0\right) \bar{e}_j =: -\frac{\partial}{\partial t} \vec{A}_s - \bar{e}_\alpha \frac{\partial}{\partial x^\alpha} A_s^0 =: \vec{E}_s. \tag{7.3}$$

The space coordinates $B_s^{\alpha\beta}$ belong to the *magnetic self-field strength*. For example is

$$B_s^{12} = -\frac{\partial}{\partial x_1} A_s^2 + \frac{\partial}{\partial x_2} A_s^1 = \frac{\partial}{\partial x^1} A_s^2 - \frac{\partial}{\partial x^2} A_s^1 = (\nabla \times \vec{A}_s) \cdot \vec{e}_3 = B_s^3. \tag{7.4}$$

With that, the *matrix representation* of the $B_s^{ij}$ reads

$$\left(B_s^{ij}\right) = \begin{pmatrix} 0 & E_s^1 & E_s^2 & E_s^3 \\ -E_s^1 & 0 & B_s^3 & -B_s^2 \\ -E_s^2 & -B_s^3 & 0 & B_s^1 \\ -E_s^3 & B_s^2 & -B_s^1 & 0 \end{pmatrix}. \tag{7.5}$$

The *mean electromagnetic self-field strength* $<\bar{\bar{B}}_s>$ follows from $\bar{\bar{B}}_s$ (7.1a) just as $<\bar{A}_s>$ (6.9) follows from $\bar{A}_s$ (6.2):

$$\begin{aligned}<\bar{\bar{B}}_s>(\bar{z}) &= iqC^4 \int_{-\infty}^{\tau} d\tau' \int_{C_r} d^4\bar{k} \, \exp[i\bar{k} \cdot (\bar{z}' - \bar{z})] \, \tilde{\rho} \, \tilde{\rho}' \, \frac{\dot{\bar{z}}'\bar{k} - \bar{k}\dot{\bar{z}}'}{\bar{k}^2} \\ &= -\partial_{\bar{z}} <\bar{A}_s> + <\bar{A}_s> \overleftarrow{\partial}_{\bar{z}}. \end{aligned} \tag{7.6}$$

The *total mean electromagnetic field strength* $<\bar{\bar{B}}>$, acting onto the particle, is the sum of the (mean) external electromagnetic field strength $<\bar{\bar{B}}_e> \approx \bar{\bar{B}}_e$ and $<\bar{\bar{B}}_s>$:

$$<\bar{\bar{B}}>(\bar{z}) = \bar{\bar{B}}_e(\bar{z}) + <\bar{\bar{B}}_s>(\bar{z}). \tag{7.7}$$

The approximation $<\bar{\bar{B}}_e> \approx \bar{\bar{B}}_e$ holds on condition that $\bar{\bar{B}}_e(\bar{z}+\bar{r})$ changes only very slowly with $\bar{r}$ for $|\bar{r}| \leq <r>$, where $<r>$ (4.9) is the particle radius. The total electromagnetic field strength $\bar{\bar{B}} = \bar{\bar{B}}_e + \bar{\bar{B}}_s$ causes the *Lorentz 4-force* (cf. [22,34])

$$\begin{aligned}\bar{F} = q <\dot{\bar{x}} \cdot \bar{\bar{B}}> &\approx q\dot{\bar{z}} \cdot <\bar{\bar{B}}> := q\bar{u} \cdot <\bar{\bar{B}}> \stackrel{(7.1c)}{=} q(u^0 \bar{e}_0 + \vec{u}) \cdot <\bar{e}_0 \vec{E} - \vec{E}\bar{e}_0 - \vec{1} \times \vec{B}> \\ &\stackrel{(6.8)}{=} q\left(\vec{u} \bullet <\vec{E}> \bar{e}_0 + u^0 <\vec{E}> + \vec{u} \times <\vec{B}>\right) \\ &=: \quad F^0 \bar{e}_0 \quad + \quad \vec{F}. \end{aligned} \tag{7.8}$$

Analogously to (6.7), we have split up the 4-speed $\dot{\bar{z}} =: \bar{u}$ into the time component $u^0 \bar{e}_0$ and the space component $\vec{u}$. The same holds for $\bar{F}$. In (7.8), $<\vec{E}>$ and $<\vec{B}>$ are the *mean electric and magnetic field strengths*, respectively.

## 8. Equation of motion

The *4-dimensional generalization of Newton's second law* reads

$$m_\infty \dot{\bar{u}} = \bar{F} = \bar{F}_e + \bar{F}_s = q\bar{u} \cdot <\bar{\bar{B}}> := q\bar{u} \cdot \left(\bar{\bar{B}}_e + <\bar{\bar{B}}_s>\right). \tag{8.1}$$

$m_\infty$ is the mechanical (bare) inertial mass. Inserting (7.6) we get

$$m_\infty \dot{\bar{u}} = q\bar{u} \cdot \bar{\bar{B}}_e + iq^2 C^4 \bar{u} \cdot \int_{-\infty}^{\tau} d\tau' \int_{C_r} d^4\bar{k} \, \exp\left[i\bar{k} \cdot (\bar{z}' - \bar{z})\right] \tilde{\rho} \tilde{\rho}' \, \frac{\bar{u}'\bar{k} - \bar{k}\bar{u}'}{\bar{k}^2}. \tag{8.2}$$

At this mean

$$C := \frac{1}{2\pi}, \quad \bar{z} = \bar{z}(\tau) = \text{4-position of the particle centre at proper time } \tau, \quad \bar{z}' := \bar{z}(\tau'),$$

$$\bar{u} := \dot{\bar{z}}(\tau) = \text{4-speed}, \quad \bar{u}' := \bar{u}(\tau'), \quad \dot{\bar{u}} = \text{4-acceleration}, \quad \bar{k} = \text{4-wave vector},$$

$$\tilde{\rho} = \tilde{\rho}(\kappa), \quad \kappa := [(\bar{k}\cdot\bar{u})^2 - \bar{k}^2]^{1/2}, \quad \tilde{\rho}' = \tilde{\rho}(\kappa'), \quad \kappa' := [(\bar{k}\cdot\bar{u}')^2 - \bar{k}^2]^{1/2},$$

$d^4\bar{k} = d\omega\, d^3\bar{k}$ = volume element in 4-$\bar{k}$-space, $C_r$ = contour in complex $\omega$-plane [see Fig. 3], $\bar{\bar{B}}_e$ = external electromagnetic 4-field strength, $q$ = charge.

In more conventional coordinate representation, the *equation of motion* (8.2) reads

$$m_\infty \dot{u}^j = q u_i B_e^{ij} + i q^2 C^4 u_i \int_{-\infty}^{\tau} d\tau' \int_{C_r} d\omega \int d^3\bar{k}\, \exp[i k_l(z'^l - z^l)]\, \tilde{\rho}\, \tilde{\rho}'\, \frac{u'^i k^j - k^i u'^j}{k_m k^m}. \tag{8.3}$$

The 4-force balance (8.2) is obviously Lorentz-invariant, because all occurring quantities are. This equation of motion is a differential-integral equation (integrodifferential equation) of second order in the solution $\bar{z}(\tau)$ sought. The solution is to be determined under the *initial conditions*

$$\bar{z}(\tau_0) := \bar{z}_0, \quad \dot{\bar{z}}(\tau_0) := \dot{\bar{z}}_0. \tag{8.4}$$

Twofold proper time integration of (8.1) yields the *equation of motion* in the form

$$\bar{z}(\tau) = \bar{z}_0 + (\tau - \tau_0)\dot{\bar{z}}_0 + \frac{q}{m_\infty} \int_{\tau_0}^{\tau} d\tau_2 \int_{\tau_0}^{\tau_2} d\tau_1\, \dot{\bar{z}}(\tau_1)\cdot <\bar{\bar{B}}>(\tau_1) \tag{8.5}$$

with $<\bar{\bar{B}}>$ corresponding to (7.7) and (7.6). This differential-integral equation implicates the initial conditions (8.4) and is suitable to derive the solution $\bar{z}(\tau)$ iteratively if the external field $\bar{\bar{B}}_e(\bar{x})$ is given. In Appendix D an example is calculated.

In Appendix C our results are compared with Nodvik´s [40].

## 9. Four self-force for very small acceleration. Zeroth approximation

We now want to calculate the *4-self-force* $\bar{F}_s := q\bar{u}\cdot<\bar{\bar{B}}_s>$. For simplicity, as a reference frame we choose the local (instantaneous) rest frame (= inertial frame) of the particle. As a reference point we choose $\bar{z}(0) = \bar{z}_0 := \bar{0}$. Let the 4-speed at this 4-point be

$$\dot{\bar{z}}(0) = \bar{u}(0) =: \bar{e}_0 \quad \text{at} \quad \bar{z}(0) = \bar{0}. \tag{9.1}$$

In this case the local rest frame coincides with the lab base $\{\bar{e}_i\}$. By means of (7.6) and (6.7), the 4-self-force gets the form [cf. (8.2)]

$$\bar{F}_s = q\bar{e}_0\cdot<\bar{\bar{B}}_s> = i q^2 C^4 \int_{-\infty}^{0} d\tau' \int d^3\bar{k} \int_{C_r} d\omega\, \exp[i(\omega z'^0 - \vec{k}\cdot\vec{z}')]\, \frac{u'^0 \vec{k} - \omega \vec{u}'}{\omega^2 - \kappa^2}\, \tilde{\rho}\, \tilde{\rho}' \tag{9.2}$$

with $\kappa := |\vec{k}|$.

If the magnitude of the 4-acceleration $\bar{a}$ is very small against the reciprocal of the particle radius (4.9), the 4-self-force $\bar{F}_s$ is in zeroth approximation proportional to $\bar{a}$:

$$\bar{F}_s \approx -m_1 \bar{a} \quad \text{for} \quad |\bar{a}| \ll <r>^{-1}. \tag{9.3}$$

The factor of proportionality $m_1$ *equals the electrostatic self-energy* (A2) of the charged particle.

To prove this claim, we calculate $\bar{F}_s$ in linear nonrelativistic approximation as to $\bar{a}$. In (9.2) the following approximations are to be inserted:

$$\vec{z}' := \vec{z}(z'^0) = \vec{z}(t') \approx \frac{\vec{a}}{2} t'^2, \quad \vec{u}' = \dot{\vec{z}}' = \vec{a} t', \quad u'^0 = \sqrt{1+\vec{u}'^2} \approx 1, \tag{9.4a}$$

$$d\tau' = dt'/u'^0 \approx dt', \quad \exp(-i\vec{k}\cdot\vec{z}') \approx 1 - i\vec{k}\cdot\vec{z}', \quad \kappa^2 = (\bar{k}\cdot\bar{e}_0)^2 - \bar{k}^2 = \vec{k}^2,$$

$$\kappa'^2 \stackrel{(6.6)}{:=} (\vec{k}\cdot\vec{u}')^2 - \vec{k}^2 = (\omega u'^0 - \vec{k}\cdot\vec{u}')^2 - \omega^2 + \vec{k}^2 \approx -2\omega\vec{k}\cdot\vec{u}' + \vec{k}^2,$$

$$\tilde{\rho}' := \tilde{\rho}(\kappa') \approx \tilde{\rho}(\kappa) + \left[\frac{d\tilde{\rho}'}{d\kappa'^2}\frac{d\kappa'^2}{d\vec{u}'}\right]_{\vec{u}'=\vec{0}} \cdot \vec{u}' = \tilde{\rho} - \frac{d\tilde{\rho}}{d\kappa^2} 2\omega\vec{k}\cdot\vec{u}', \qquad (9.4b)$$

$$\tilde{\rho}\tilde{\rho}' \approx \left(1 - \omega\vec{k}\cdot\vec{u}'\frac{d}{d\kappa^2}\right)\tilde{\rho}^2.$$

Hence, (9.2) reads

$$\vec{F}_s \approx iq^2 C^4 \int_{-\infty}^{0} dt' \int d^3\vec{k} \int_{C_r} d\omega\, e^{i\omega t'}(1-i\vec{k}\cdot\vec{z}')\frac{\vec{k}-\omega\vec{u}'}{\omega^2-\kappa^2}\left(1-\omega\vec{k}\cdot\vec{u}'\frac{d}{d\kappa^2}\right)\tilde{\rho}^2$$

$$\approx -iq^2 C^4 \int_{-\infty}^{0} dt' \int d^3\vec{k} \int_{C_r} d\omega\, e^{i\omega t'} \frac{1}{\omega^2-\kappa^2}\left\{\omega\left[\vec{u}'+\vec{k}\vec{k}\cdot\vec{u}'\frac{d}{d\kappa^2}\right] + i\vec{k}\vec{k}\cdot\vec{z}'\right\}\tilde{\rho}^2. \qquad (9.4c)$$

In the integrand we have omitted the $\vec{k}$-odd term, since for symmetry reasons

$$\int d^3\vec{k}\,\frac{\vec{k}\tilde{\rho}^2}{\omega^2-\kappa^2} = \vec{0}.$$

The volume element in $\vec{k}$-space may be factorized:

$$d^3\vec{k} = d\kappa\,\kappa^2\,d\Omega,$$

where $d\Omega$ is the solid angle element. The directional mean of the dyadic product $\vec{k}\vec{k}$ is

$$\frac{1}{4\pi}\int d\Omega\,\vec{k}\vec{k} =: \langle\vec{k}\vec{k}\rangle_\Omega = N\kappa^2\vec{1} := N\kappa^2\vec{e}_\alpha\vec{e}_\alpha,$$

where $\vec{1}$ means the 3-unit dyadic. The last but one of these equalities holds for symmetry reasons. The number $N$ can easily be determined by contraction:

$$\langle\vec{k}\cdot\vec{k}\rangle_\Omega = \kappa^2 = N\kappa^2 3 \Rightarrow \langle\vec{k}\vec{k}\rangle_\Omega = \frac{\kappa^2}{3}\vec{1}. \qquad (9.5)$$

With that, (9.4c) goes over into

$$\vec{F}_s \approx -iq^2 C^3 \int_{-\infty}^{0} dt' \int_{-\infty}^{\infty} d\kappa\,\kappa^2 \int_{C_r} d\omega\,\frac{e^{i\omega t'}}{\omega^2-\kappa^2}\left\{\omega\left[\vec{u}'+\frac{\kappa^2}{3}\vec{u}'\frac{d}{d\kappa^2}\right] + \frac{i}{3}\kappa^2\vec{z}'\right\}\tilde{\rho}^2. \qquad (9.6)$$

Performing the $\omega$-integration with the calculus of residues [33] yields

$$\vec{F}_s \approx -q^2 C^2 \int_{-\infty}^{0} dt' \int_{-\infty}^{\infty} d\kappa\,\kappa^2\left\{\cos\kappa t'\left[\vec{u}'+\frac{\kappa}{6}\vec{u}'\frac{d}{d\kappa}\right] - \frac{\kappa}{3}\vec{z}'\sin\kappa t'\right\}\tilde{\rho}^2. \qquad (9.7)$$

The time integral with $\vec{z}'$ can be transformed by partial integration:

$$\int_{-\infty}^{0} dt'\,\vec{z}'\sin\kappa t' = -\left[\vec{z}(t')\frac{\cos\kappa t'}{\kappa}\right]_{-\infty}^{0} + \int_{-\infty}^{0} dt'\,\vec{u}'\frac{\cos\kappa t'}{\kappa}.$$

The term outside the integral vanishes. Consequently, (9.7) reads with $d/d\kappa = -\overleftarrow{d}/d\kappa$ (partial integration)

$$\vec{F}_s = -q^2 C^2 \int_{-\infty}^{0} dt'\,\vec{u}' \int_{-\infty}^{\infty} d\kappa\,\kappa^2 e^{i\kappa t'}\left(\frac{2}{3} - \frac{\kappa}{6}\frac{\overleftarrow{d}}{d\kappa}\right)\tilde{\rho}^2\left[1+O(\vec{u}'^2)\right]$$

$$\approx -\frac{q^2}{24\pi^2}\int_{-\infty}^{0} dt'\,\vec{u}' \int_{-\infty}^{\infty} d\kappa\,(\kappa^2 - it'\kappa^3)e^{i\kappa t'}\tilde{\rho}^2. \qquad (9.8)$$

Inserting

$$\vec{u}' \approx \vec{a}\,t' \qquad (9.9)$$

and integrating over $t'$, we get

$$\bar{F}_s \approx -\bar{a}\frac{q^2}{8\pi^2}\int_{-\infty}^{\infty} d\kappa\,\tilde{\rho}^2 \stackrel{(A2)}{=} -\bar{a}\,m_1, \qquad \text{q.e.d.} \tag{9.10}$$

If we had approximated $\tilde{\rho}'$ not by (9.4b), but by $\tilde{\rho}$, the result would have been $\bar{F}_s \approx -(4/3)m_1\bar{a}$. *The correction term $\tilde{\rho}' - \tilde{\rho}$ in (9.4b) just eliminates the notorious factor 4/3*, which violates the relativistic invariance [28,54,59]. The result (9.10) coincides with that of Kolbenstvedt [32], who derived it (without Fourier transformation) for a constant linear acceleration.

By proper time integration of the 4-self-force (9.3) one gets (approximately for small acceleration) the 4-momentum of the self-field:

$$-\int d\tau\,\bar{F}_s \approx \bar{p}_s = m_1\bar{u}, \quad |\bar{a}| \ll <r>^{-1}. \tag{9.10a}$$

The last equation has already been derived by Fermi [16] and Rohrlich [53], yet only for a constant 4-velocity $\bar{u}$. They calculated $\bar{p}_s$ as a Lorentz invariant volume integral over the 4-momentum density of the electromagnetic self-field at constant proper time $\tau$.

If we replace in (9.2)

$$\tilde{\rho}' \to \frac{3}{4}\tilde{\rho}, \tag{9.11}$$

we get the correct result (9.10).

With (9.10), the *equation of motion* (8.2) can *in zeroth approximation* be written

$$m_0\bar{a} \approx \bar{F}_e, \quad \text{where} \quad m_0 = m_\infty + m_1 \tag{9.12}$$

means the (observable) inertial mass for small acceleration of the particle.

For the structure function $\tilde{\rho}_1(\kappa) = (1 + r_1^2\kappa^2)^{-1/2}$ (4.6), the *electrostatic mass* becomes

$$m_1 \stackrel{(A2)}{:=} \frac{\mu_0 q^2}{8\pi^2}\int_{-\infty}^{\infty} d\kappa\,\tilde{\rho}^2 = \frac{\mu_0 q^2}{8\pi r_1}. \tag{9.13}$$

We have completed the right side with the magnetic field constant $\mu_0 := (\varepsilon_0 c^2)^{-1}$. With (4.7) we get for an electron (or any other simply charged particle)

$$m_1 = \alpha\,m_0, \quad \text{where} \quad \alpha := \frac{q^2}{4\pi\varepsilon_0 c\hbar} \approx \frac{1}{137} \tag{9.14}$$

is the fine structure constant.

## 10. Four-self-force in second approximation

If we insert the replacement (9.11) in (9.2), we get a *second approximation for the 4-self-force*:

$$\bar{F}_s(0) = \bar{F}_s = \frac{3}{4}iq^2C^4\int_{-\infty}^{\infty} d\tau'\int d^3\vec{k}\,\tilde{\rho}^2(\kappa)\int_{C_r} d\omega\,\exp\left[i\left(\omega z'^0 - \vec{k}\cdot\vec{z}'\right)\right]\frac{u'^0\vec{k} - \omega\vec{u}'}{\omega^2 - \kappa^2}. \tag{10.1}$$

The $\omega$-integration yields, using the method of residues [33, 38],

$$\bar{F}_s = \frac{3}{4}q^2C^3\int d^3\vec{k}\,\tilde{\rho}^2\int_{-\infty}^{0} d\tau'\,\exp(-i\vec{k}\cdot\vec{z}')\left(\frac{iu'^0\vec{k}}{\kappa}\sin\kappa z'^0 - \vec{u}'\cos\kappa z'^0\right). \tag{10.2}$$

Integration by parts over the retarded proper time $\tau'$ delivers

$$\vec{F}_s = \frac{3}{4}q^2 C^3 \int d^3\vec{k}\, \tilde{\rho}^2 \left\{ \frac{\vec{k}}{i\kappa^2}\left[1 - \exp\left(-i\vec{k}\cdot\vec{z}(-\infty)\right)\cos\left(\kappa z^0(-\infty)\right)\right] \right.$$
$$\left. - \int_{-\infty}^{0} d\tau' \exp\left(-i\vec{k}\cdot\vec{z}'\right) \vec{P}_{\perp\vec{k}}\cdot\vec{u}'\, \cos\kappa z'^0 \right\}, \tag{10.3}$$

where the projection dyadic

$$\vec{P}_{\perp\vec{k}} := \vec{1} - \frac{\vec{k}\vec{k}}{\kappa^2} =: \vec{1} - \vec{e}_{\vec{k}}\vec{e}_{\vec{k}} \tag{10.4}$$

projects a space vector (here $\vec{u}'$) onto a plane perpendicular to the wave vector $\vec{k}$. The term proportional to $\vec{k}$ can be omitted, because both $\vec{k}$-space integrals (with $\tau' = 0$ and $\tau' = -\infty$) vanish. (The former integral vanishes, because the integrand is odd in $\vec{k}$; the latter, because the integrand oscillates very rapidly for $\tau' \to -\infty$.) With $d\tau'\vec{u}' = d\vec{z}'$ the *self-force* simplifies to

$$\vec{F}_s = -\frac{3}{4}q^2 C^3 \int d^3\vec{k}\, \tilde{\rho}^2 \int_{\vec{z}'(-\infty)}^{\vec{0}} d\vec{z}'\cdot\vec{P}_{\perp\vec{k}}\, \cos\kappa z'^0 \cos\vec{k}\cdot\vec{z}' \ . \tag{10.5}$$

This formula differs from nonrelativistic derivations [8,21,47] by the factor 3/4.

The $\vec{k}$-space integral can be calculated by introduction of spherical coordinates:

$$d^3\vec{k} = \kappa^2 d\kappa\, d\Omega, \quad d\Omega = \sin\vartheta\, d\vartheta\, d\varphi, \quad \vartheta = \angle(\vec{k},\vec{z}'),$$

$$\vec{F}_s = -\frac{3}{2}q^2 C^2 \int_{\vec{z}'(-\infty)}^{\vec{0}} d\vec{z}'\cdot\int_0^{\infty} d\kappa\, \kappa^2 \tilde{\rho}^2 \cos\kappa z'^0 \frac{1}{4\pi}\int d\Omega \left(\vec{1} - \vec{e}_{\vec{k}}\vec{e}_{\vec{k}}\right)\cos\vec{k}\cdot\vec{z}' . \tag{10.6}$$

With

$$\vec{e}_{\vec{k}} := \frac{\vec{k}}{\kappa} = \sin\vartheta\cos\varphi\, \vec{e}_1' + \sin\vartheta\sin\varphi\, \vec{e}_2' + \cos\vartheta\, \vec{e}_3', \quad \vec{e}_3' := \frac{\vec{z}'}{|\vec{z}'|}, \tag{10.7}$$

the last solid angle integral can easily be evaluated:

$$M_1 := \frac{1}{4\pi}\int d\Omega\, \cos\vec{k}\cdot\vec{z}' =: \langle\cos\vec{k}\cdot\vec{z}'\rangle$$
$$= \frac{1}{2}\int_0^\pi d\vartheta\, \sin\vartheta\, \cos(\kappa|\vec{z}'|\cos\vartheta) = \frac{\sin\lambda}{\lambda}, \qquad \lambda := \kappa|\vec{z}'|; \tag{10.8}$$

$$\vec{M}_2 := \langle\vec{e}_{\vec{k}}\vec{e}_{\vec{k}}\cos\vec{k}\cdot\vec{z}'\rangle = \frac{1}{2}(1+d_\lambda^2)M_1 \vec{P}_{\perp\vec{z}'} - d_\lambda^2 M_1 \vec{P}_{\vec{z}'}, \quad d_\lambda := \frac{d}{d\lambda}; \tag{10.9}$$

$$\vec{P}_{\vec{z}'} := \vec{e}_3'\vec{e}_3' = \vec{1} - \vec{e}_1'\vec{e}_1' - \vec{e}_2'\vec{e}_2' = \vec{1} - \vec{P}_{\perp\vec{z}'}\ ; \tag{10.10}$$

$$\vec{M}(\lambda) = \vec{M}(-\lambda) = M_1\vec{1} - \vec{M}_2$$
$$= \frac{2}{\lambda^3}(\sin\lambda - \lambda\cos\lambda)\vec{P}_{\vec{z}'} + \frac{1}{\lambda^3}\left[(\lambda^2-1)\sin\lambda + \lambda\cos\lambda\right]\vec{P}_{\perp\vec{z}'}$$
$$=: M_\parallel \vec{P}_{\vec{z}'} + M_\perp \vec{P}_{\perp\vec{z}'}$$
$$= 4\left(\frac{1}{3!} - \frac{2}{5!}\lambda^2 + \frac{3}{7!}\lambda^4 - \cdots\right)\vec{P}_{\vec{z}'} + 4\left(\frac{1}{3!} - \frac{2^2}{5!}\lambda^2 + \frac{3^2}{7!}\lambda^4 - \cdots\right)\vec{P}_{\perp\vec{z}'}$$
$$= \frac{2}{3}\left(\vec{1} - \frac{1}{10}\lambda^2\vec{P}_{\vec{z}'} - \frac{1}{5}\lambda^2\vec{P}_{\perp\vec{z}'}\right) + O(\lambda^4). \tag{10.11}$$

With that, the *self-force* at $\vec{z} = \vec{0}$ reads

$$\vec{F}_s = -\frac{3q^2}{16\pi^2} \int_{-\infty}^{\infty} d\kappa\, \kappa^2 \tilde{\rho}^2 \int_{\vec{z}'(-\infty)}^{\vec{0}} d\vec{z}' \cdot \vec{\vec{M}}(\kappa\,|\,\vec{z}'|) \cos\kappa z'^0. \tag{10.12}$$

Inserting the structure function $\tilde{\rho}_1(\kappa)$ (4.6) and calculating the $\kappa$-integral with the residues theorem one gets

$$\vec{F}_s = \frac{m_1}{r_1^2} \int_{\vec{z}'(-\infty)}^{\vec{0}} d\vec{z}' \cdot \vec{\vec{A}}\left(\frac{|\vec{z}'|}{r_1}\right) \exp\frac{t'}{r_1}, \tag{10.13}$$

with the electrostatic mass $m_1$ (9.14) and the dyadic

$$\vec{\vec{A}}(\alpha) := \frac{3}{2}\vec{\vec{M}}(i\alpha) =: \frac{\vec{z}'\vec{z}'}{|\vec{z}'|^2} A_\|(\alpha) + \left(\vec{\vec{1}} - \frac{\vec{z}'\vec{z}'}{|\vec{z}'|^2}\right) A_\perp(\alpha), \tag{10.14}$$

$$A_\|(\alpha) := \frac{3}{2} M_\|(i\alpha) = \frac{3}{\alpha^3}(\alpha\cosh\alpha - \sinh\alpha)$$

$$= 1 + \frac{\alpha^2}{10} + \frac{\alpha^4}{280} + O(\alpha^6)$$

$$= 6 \sum_0^\infty \frac{1+n}{(3+2n)!} \alpha^{2n}, \tag{10.15}$$

$$A_\perp(\alpha) := \frac{3}{2} M_\perp(i\alpha) = \frac{3}{2\alpha^3}\left[(1+\alpha^2)\sinh\alpha - \cosh\alpha\right]$$

$$= 1 + \frac{\alpha^2}{5} + \frac{3\alpha^4}{280} + O(\alpha^6)$$

$$= 6 \sum_0^\infty \frac{(1+n)^2}{(3+2n)!} \alpha^{2n}. \tag{10.16}$$

With $d\vec{z}' = \dot{\vec{z}}\,d\tau'$, the self-force at $\vec{z} = \vec{0}$ can also be written in the form

$$\vec{F}_s = \frac{m_1}{r_1^2} \int_{-\infty}^0 d\tau' \left\{ \frac{\dot{\vec{z}}'\cdot\vec{z}'}{|\vec{z}'|^2} \vec{z}'\left[A_\|\!\left(\frac{|\vec{z}'|}{r_1}\right) - A_\perp\!\left(\frac{|\vec{z}'|}{r_1}\right)\right] + \dot{\vec{z}}' A_\perp\!\left(\frac{|\vec{z}'|}{r_1}\right) \right\} \exp\frac{t'(\tau')}{r_1}. \tag{10.17}$$

*The choice of the reference point* $\vec{z}(0) = \vec{0}$ *does not mean a restriction of generality*, as we can arbitrarily choose the zero of the proper time scale. In each case, $\vec{z}' := \vec{z}(\tau')$ means the spatial (vectorial) distance of $\vec{z}(\tau')$ [as to the local rest frame $\{\vec{e}_i'\}$ (2.4); cf. (6.7)] from the reference point on the worldline. The vector $\vec{z}'$ lies in the 3-plane $\sigma$ [see (2.9) and Fig. 1] through the reference point.

## 11. Rectilinear motion

For a *rectilinear motion* in $x$-direction, the 4-position vector of the particle's centre specializes to the *2-position vector*

$$\bar{z} = t\bar{e}_0 + \vec{z} = t\bar{e}_0 + x\bar{e}_1, \qquad x_2 = x_3 = 0. \tag{11.1}$$

Analogously holds for the *external 2-force*

$$\bar{F}_e = F_e \bar{w} \quad \text{with} \quad \bar{w} \stackrel{(2.3)}{:=} u^1 \bar{e}_0 + u^0 \bar{e}_1 =: \frac{\bar{a}}{a}, \quad \bar{w}\cdot\bar{u} = 0 . \tag{11.2}$$

Here $\bar{w}$ means the spacelike unit vector parallel to the *2-acceleration* $\bar{a} = a\bar{w}$, which is orthogonal to the 2-speed

$$\bar{u} := \dot{\bar{z}} = u^0 \bar{e}_0 + u^1 \bar{e}_1 .$$

We now calculate the *self-force* (10.13) at $\vec{z} = x\vec{e}_1 = \vec{0}$. With

$$\bar{w} = \bar{e}_1, \quad \bar{F}_s = F_s \bar{e}_1, \quad \vec{z}' = x'\vec{e}_1, \quad d\vec{z}'\cdot\vec{\tilde{A}}\left(\frac{|\vec{z}'|}{r_1}\right) = dx' A_{\parallel}\left(\frac{x'}{r_1}\right)\bar{e}_1, \tag{11.3}$$

eq. (10.13) specializes for a rectilinear motion to

$$F_s = \frac{m_1}{r_1^2} \int_{x'(-\infty)}^{0} dx' \exp\frac{t'(x')}{r_1} A_{\parallel}\left(\frac{x'}{r_1}\right) . \tag{11.4}$$

The integration runs over the past $t' = -\infty \ldots 0$. At time $t = 0$, the *initial conditions* (9.1) shall hold:

$$x(0) = \frac{dx}{dt}(0) = 0. \tag{11.5}$$

Then, for

$$F_e > 0, \quad t' < 0 \quad \text{we have} \quad x'(-\infty) = \infty, \quad dx' < 0. \tag{11.6}$$

That is, as the integrand in (11.4) is positive,

$$F_s F_e < 0: \tag{11.7}$$

*The self-force is oppositely directed to the external force* [cf. Fig. 5 and (13.2)].

From (11.2) and (8.1) follows that $\bar{F}_e$, $\bar{F}_s$ and $\bar{a}$ are parallel 2-vectors. This can also be expressed by the equations

$$\bar{F}_e = m\bar{a}, \quad \bar{F}_s = -m_s \bar{a}. \tag{11.8a,b}$$

We call $m$ *effective inertial mass* and $m_s$ *(effective) electrodynamic mass*. They are connected via $m_\infty \bar{a} = \bar{F}_e + \bar{F}_s$ (8.1):

$$m = m_\infty + m_s . \tag{11.9}$$

*Approximately holds*

$$m \approx m_0 = m_\infty + m_1 \quad \text{for} \quad |\bar{a}| \stackrel{(4.9)}{\ll} <r>^{-1}. \tag{11.10}$$

## 12. First example: Rectilinear motion in a homogeneous electric field

For a constant external electric field $E_e = F_e / q$ *the position-time law*

$$x = \frac{1}{f}\left(\sqrt{1 + f^2 t^2} - 1\right) \quad \text{with} \quad f = \frac{F_e}{m} \tag{12.1}$$

**Fig. 4**  satisfies, as you know, the equation of motion (11.8a) (with time-independent $m$ ) and the initial conditions (11.5); see Fig. 4 and [5,22,25,54]. The *hyperbolic worldline* (12.1) may also be derived by iteration of (8.5); see App. D.

The *(effective) field strength* $f$ depends on the *asymptotic field strength*

$$f_\infty := F_e / m_\infty \ . \tag{12.2}$$

To determine the dependence of $f$ on $f_\infty$, we insert the inverse function of (12.1),

$$t = \text{sign}\, t \sqrt{x(x + 2/f)} \ , \tag{12.3}$$

into (11.4):

$$F_s \;=\; \frac{m_1}{r_1^2} \int_{(\text{sgn}\, f)\infty}^{0} dx' \exp\!\left[-\frac{1}{r_1}\sqrt{x'(x'+2f^{-1})}\right] A_\parallel\!\left(\frac{x'}{r_1}\right) . \tag{12.4}$$

Introducing this *integral representation of the self-force* into (8.1) yields the connection between the field strengths $f$ and $f_\infty$ at $\overline{z} = \overline{0}$:

$$f \;=\; f_\infty - \frac{r_\infty}{r_1^3} \int_{0}^{(\text{sgn}\, f)\infty} dx' \exp\!\left[-\frac{1}{r_1}\sqrt{x'(x'+2f^{-1})}\right] A_\parallel\!\left(\frac{x'}{r_1}\right) =: f_\infty + f_s \ , \tag{12.5}$$

$$r_\infty := \frac{q^2}{8\pi m_\infty}, \quad f := \frac{qE_e}{m}, \quad f_\infty := \frac{qE_e}{m_\infty}, \quad f_s := \frac{F_s}{m_\infty} := \frac{q\langle E_s \rangle}{m_\infty}. \tag{12.6}$$

**Fig. 5**  From it we can determine $f_\infty(f)$ or $f(f_\infty)$ (see Fig. 5 and App. B).

To simplify the notation, let us introduce dimensionless variables:

$$\delta_\infty := r_1 f_\infty = \frac{r_1 q E_e}{m_\infty}, \quad \delta := r_1 f = r_1 \ddot{x}(0) = r_1 a, \quad \alpha = \frac{x'}{r_1}, \tag{12.7}$$

$$\Delta(\delta) := \int_{0}^{(\text{sgn}\,\delta)\infty} d\alpha \exp\!\left[-\sqrt{\alpha^2 + 2\alpha/\delta}\right] \frac{3}{\alpha^3}(\alpha \cosh\alpha - \sinh\alpha) \;=\; -\frac{r_1}{m_1} F_s \ . \tag{12.8}$$

With that, (12.5) reads

$$\delta_\infty = \delta + \frac{m_1}{m_\infty} \Delta(\delta) \ . \tag{12.9}$$

The integral (12.8) was numerically integrated (with Simpson's rule; see Table 1 and App. E). *Asymptotic approximations* are

$$\Delta(\delta) \approx \delta \quad \text{for} \quad 0 \le \delta \ll 1, \tag{12.10a}$$

$$\simeq 1.5 \quad \text{for} \quad \delta \gg 1. \tag{12.10b}$$

$\Delta(\delta)$ is an odd function:

$$\Delta(-\delta) := -\Delta(\delta). \tag{12.11}$$

This function and the *relative reactive mass*

$$\mu_r(\delta) := 1 - \frac{\Delta(\delta)}{\delta} = \frac{F_r}{m_1 f} =: \frac{m_r}{m_1} \tag{12.12}$$

are not analytic at $\delta = r_1 f = 0$. Therefore, the self-force $F_s[r_1]$ (12.8) and the *reaction force*

$$F_r[r_1] := m_1 a + F_s \tag{12.13}$$

are not developable into Laurent series at $r_1 = 0$. We suppose that this statement is not only valid for our special structure function (4.6).

A good *approximation for* $\mu_r(\delta)$ is

$$\tilde{\mu}_r(\delta) := \exp\left(\frac{-1.5}{\sqrt{|\delta|}\left(0.1078 + \sqrt{|\delta|}\right)}\right), \qquad (12.14)$$

see Table 1.

Corresponding to (12.7), (12.8) and (12.12), the *self-force* $F_s$ (12.4) may be written in the forms

$$F_s = -\frac{m_1}{r_1}\Delta(r_1 a) = -m_1 a[1-\mu_r(r_1 a)]. \qquad (12.15)$$

A formal Taylor expansion in powers of the acceleration $a$ yields

$$F_s(a) = -m_1 a + 0 \cdot a^2 + 0 \cdot a^3 + \cdots . \qquad (12.16)$$

This asymptotic expansion at $a = 0$, where $F_s(a)$ is nonanalytic, also holds for the exact self-force (9.2) and for the approximated form with $\tilde{\mu}_r(r_1 a)$ (12.14) instead of $\mu_r(r_1 a)$ (12.12).

For large $a$, the self-force $F_s$ (12.15) remains finite:

$$F_s \simeq -1.5 \frac{m_1}{r_1} \quad \text{for} \quad a \gg <r>^{-1}, \qquad (12.17)$$

as it follows from (12.8) and (12.10b). Hence, the *equation of motion asymptotically reads*

$$m_\infty a = F_e + F_s \simeq F_e \quad \text{for} \quad a \gg <r>^{-1}. \qquad (12.18)$$

The finiteness of $F_s$ (12.17) and the deduction (12.18) also hold for other rectilinear motions in extreme electric fields. For, in this case the particle asymptotically moves with light velocity. That is, in (11.4) we have to insert $t' \simeq x'$, which yields (12.17), too.

These asymptotic relations have more mathematical than physical interest, since $a \gg <r>^{-1}$ passes over the restriction $a \ll <r>^{-1}$ (3.6). In this case of extreme acceleration we would have to take into account the nonapproximated functional determinant $D = 1 - \bar{r} \cdot \bar{a}$ (3.3).

## 13. Connections between forces, masses and field strengths

From Fig. 5 is evident: The *reaction force*

$$F_r := m_0 a - F_e = m_1 a + F_s =: m_0 f_r \qquad (13.1)$$

has the direction of the external force $F_e =: mf = ma$. The *self-force* $F_s =: m_\infty f_s = -m_s f$ has the opposite direction [see (11.7)]. These forces are connected as follows:

$$F_e = m_\infty a - F_s = m_0 a - F_r = ma = -\frac{m}{m_s}F_s = \frac{m}{m_r}F_r. \qquad (13.2)$$

Between the positive definite masses $m_\infty, m_1, m_0, m, m_r, m_s$ hold the relations

$$m_\infty + m_1 = m_0 = \lim_{|a|\to 0} m(|a|) > m = m_0 - m_r = m_\infty + m_s > m_\infty = \lim_{|a|\to\infty} m(|a|). \qquad (13.3)$$

With $f_0 = F_e/m_0$ instead of $f_\infty = F_e/m_\infty$ and with the relative reactive mass $\mu_r =: m_r/m_1$ (12.12), equation (12.9) may also be written in the form

$$f_0 = f\left[1 - \mu_1 \mu_r\left(\frac{f r_0}{\mu_1}\right)\right], \quad \mu_1 := \frac{m_1}{m_0}, \quad r_0 := \frac{q^2}{8\pi m_0}. \qquad (13.4)$$

**Fig. 6**    This function $f_0(f)$ is visualized in Fig. 6.

As long as the *relative electrodynamic mass* $\mu_1$ is smaller than 1, the acceleration $a = f := F_e / m$ tends with the field strength $f_0 := F_e / m_0$ to infinity. For $\mu_1 = 1$ this happens already at $f_0 = 1.5 / r_0$. This unphysical behaviour is only to exclude if the inequality

$$\mu_1 = \frac{m_1}{m_0} = \frac{r_0}{r_1} < 1 \tag{13.5}$$

holds. That is, $r_1$ must be larger than $r_0$. Hence, the *hypothesis* $m_1 = m_0$ *of Abraham and Lorentz* [1,35] *and electric point charges* ($r_1 = 0$ [12,43,54]) *are excluded*.

## 14. Effective inertial mass

In analogy to $m_\infty = qE_e / f_\infty = F_e / f_\infty$ and $m_0 = F_e / f_0$, we have already defined in (12.1) by means of the effective field strength $f$ an *effective inertial mass*

$$m := \frac{F_e}{f} = \frac{qE_e}{f}. \tag{14.1}$$

Dividing eq. (13.4) by $f$, we get the *relative effective mass*

$$\mu := \frac{m}{m_0} = \frac{f_0}{f} = 1 - \mu_1 \mu_r \left( \frac{r_0 f}{\mu_1} \right). \tag{14.2}$$

**Fig. 7**  In Fig. 7 it is displayed as a function of $\delta_0 = r_0 f = r_0 \ddot{x}(0) = r_0 a$ with the relative electrodynamic mass $\mu_1 := m_1 / m_0$ as a parameter. This function $\mu(\delta_0)$ decreases from 1 to $\mu_\infty = 1 - \mu_1$ if $|\delta_0|$ increases from 0 to $\infty$. Hence, the effective mass $m = \mu m_0$ decreases from $m_0$ to $m_\infty$ with increasing magnitude of acceleration $|a| = |\delta_0| / r_0$:

$$m_0 > m > m_\infty \quad \text{for} \quad 0 < |a| < \infty. \tag{14.3}$$

With $f = a$, eq. (14.1) may be written in the form of *Newton's second law*:

$$\frac{\mu m_0 a}{q E_e} = 1. \tag{14.4}$$

It is a matter of discretion, whether one combines $\mu$ with $m_0$ to the *effective inertial mass* $m := \mu m_0$ (14.2) or with the acceleration $a$ to the *effective acceleration*

$$a_{eff} := \mu a \tag{14.5}$$

or with $E_e$ to the *effective external electric field strength*

$$E_{e,eff} := \mu^{-1} E_e = \frac{m_0}{m_\infty} (E_e + <E_s>) \tag{14.6}$$

or with the external force $F_e = q E_e$ to the *effective external force*

$$F_{e,eff} := \mu^{-1} F_e \tag{14.7}$$

or with the charge $q$ to the *effective charge*

$$q_{eff} := \mu^{-1} q. \tag{14.8}$$

A *test for the quality of the classical theory* exposed would be to check the equation of motion

$$m \bar{a} = \bar{F}_e \tag{14.9}$$

experimentally. Then principally from the run of $\mu(\delta_0)$ (14.2), the relative electrodynamic mass $\mu_1 := m_1/m_0$ could be determined. Certainly, for extreme electric field strengths $E_e$, quantum mechanical effects superpose themselves [61].

To estimate the possibility of an experimental test of the equation of motion (14.9), we calculate the ratio $F_r/F_e =$ reaction force/external force. By means of (12.12) and (14.1), we get

$$\frac{F_r}{F_e} = \frac{m_1 \mu_r(\delta)}{m} . \qquad (14.10)$$

For

$$\delta \stackrel{(12.7)}{:=} r_1 F_e / mc^2 = 0.1 \qquad (14.11)$$

the ratio (14.10) becomes $F_r/F_e \approx 10^{-5} \mu_1$. Assuming $r_1 = \lambdabar_C/2$ (4.7) and $m \approx m_0$, one gets from (14.11) the *critical electric field strength* (cf. [35])

$$E_{0.1} = \frac{F_e}{q} := \frac{0.2 m_0 c^2}{q \lambdabar_C} = \frac{0.2 m_0^2 c^3}{q \hbar} . \qquad (14.12)$$

For a positron it becomes $E_{0.1} = 2.6 \times 10^{17}$ V m$^{-1}$. The corresponding critical acceleration is

$$a_{0.1} = \frac{q E_{0.1}}{m_0} = 4.6 \times 10^{28} \, m\, s^{-2} . \qquad (14.13)$$

Below this gigantic electric field strength and acceleration, the reaction force $F_r$ is practically always negligible against $F_e = qE_e$. Therefore, in case of a homogeneous electric field, our calculations have mainly theoretical interest in the consistency of classical electrodynamics. The equation of motion $m_0 \bar{a} \approx \bar{F}_e$ is a very good approximation at $|\bar{F}_e| = const.$

A *better approximation* than $\bar{a} = \bar{f} := \bar{F}_e/m \approx \bar{f}_0 := \bar{F}_e/m_0$ is [cf. (14.10) and (12.14)]

$$\bar{a} = \bar{f} =: \bar{f}_0 + \bar{f}_r \approx \left[1 + \mu_1 \exp\left(\frac{-14c}{\sqrt{r_1 |\bar{f}_0|}}\right)\right] \bar{f}_0, \qquad |\bar{f}_0| \ll \frac{c^2}{r_1} . \qquad (14.14)$$

This relativistically invariant equation of rectilinear motion also approximately holds for external field strengths $\bar{f}_0 := q\dot{\bar{z}} \cdot \bar{\bar{B}}_e(\bar{z})/m_0$, slowly depending on the particle's 2-position $\bar{z} = t\bar{e}_0 + x\bar{e}_1$ and its proper time derivative $\dot{\bar{z}}$. At given $\bar{f}_0(\bar{z}, \dot{\bar{z}})$, eq. (14.14) is a *differential equation of second order for* $\bar{z}(\tau)$. Its solution will in general more easily to be found than that of the corresponding exacter differential-integral equation (8.2).

## 15. Four-self-force in first approximation

Corresponding to (9.4a), (10.8) and (10.11) holds for $|\bar{a}| \ll <r>^{-1}$:

$$\bar{z}' \approx \frac{\bar{a}}{2} t'^2, \quad \cos \vec{k} \cdot \vec{z}' \approx 1, \quad \vec{M}(\lambda) \approx \vec{M}\left(\kappa \frac{|\bar{a}|}{2} t'^2\right) \approx \frac{2}{3} \vec{1} . \qquad (15.1)$$

With that, the second approximation (10.12) goes over into the less accurate *first approximation for the self-force* at $\bar{z} = t\bar{e}_0 + \vec{z} = \bar{0}$:

$$\bar{F}_s = \vec{F}_s = -\frac{q^2}{8\pi^2} \int_{-\infty}^{0} dt' \frac{d\vec{z}'}{dt'} \int_{-\infty}^{\infty} d\kappa \, \kappa^2 \cos \kappa t' \, \tilde{\rho}^2 . \qquad (15.2)$$

Twofold partial integration of the time integral gives

$$\overline{F}_s \approx -\frac{q^2}{8\pi^2} \int_{-\infty}^{\infty} d\kappa \, \tilde{\rho}^2 \left( \ddot{\vec{z}}(0) - \int_{-\infty}^{0} dt' \frac{d^3\vec{z}'}{dt'^3} \cos\kappa t' \right)$$

$$\stackrel{(9.14)}{=} -m_1 \ddot{\vec{z}}(0) + \int_{-\infty}^{0} dt' \frac{d^3\vec{z}'}{dt'^3} Q(t'), \quad \ddot{\vec{z}}(0) := \frac{d^2\vec{z}}{d\tau^2}(0) = \frac{d^2\vec{z}}{dt^2}(0). \tag{15.3}$$

The function

$$Q(t) := \frac{q^2}{8\pi^2} \int_{-\infty}^{\infty} d\kappa \, \tilde{\rho}^2 \cos\kappa t \quad \text{with} \quad Q(0) \stackrel{(A2)}{=} m_1 \tag{15.4}$$

is called *memory function* [20]. For our special structure function $\tilde{\rho}_1 = (1+r_1^2\kappa^2)^{-1/2}$ (4.6), we get

$$Q_1(t) = m_1 \exp\left(-\frac{|t|}{r_1}\right). \tag{15.5}$$

With that, the first approximation (15.2) of the 4-self-force at $\tau = 0$ reads

$$\overline{F}_s \approx -m_1 \ddot{\vec{z}}(0) + m_1 \int_{-\infty}^{0} dt' \frac{d^3\vec{z}(t')}{dt'^3} \exp\frac{-|t'|}{r_1}. \tag{15.6}$$

## 16. Second example: Spiraling of a charged particle in a homogeneous magnetic field

As another example for the (approximate) solution of the equation of motion (8.2), we consider the *plane motion of a charged particle in a homogeneous magnetic field*. The constant external magnetic field strength (induction) may have the negative 3-direction:

$$\vec{B}_e = -B_e \vec{e}_3. \tag{16.1}$$

To simplify the calculation, we first decompose [analogously to (6.7)] the vectors $\overline{z}, \overline{u}, \dot{\overline{u}}, \overline{F}_e$ and $\overline{F}_s$, occurring in the equation of motion (8.2), into *time and space components*:

$$\overline{z} = z^0 \overline{e}_0 + \vec{z}, \quad \dot{\overline{z}}(\tau) =: \overline{u} = u^0 \overline{e}_0 + \vec{u}, \quad \dot{\overline{u}} = \dot{u}^0 \overline{e}_0 + \dot{\vec{u}}, \tag{16.2a}$$

$$\overline{F}_e \stackrel{(7.8)}{=} q\vec{u} \times \vec{B}_e = \vec{F}_e, \quad \overline{F}_s = F_s^0 \overline{e}_0 + \vec{F}_s. \tag{16.2b}$$

With that, the *equation of motion* (8.2) splits up into a *time* and a *space* part:

$$m_\infty \dot{u}^0 = F_s^0, \tag{16.3a}$$

$$m_\infty \dot{\vec{u}} = q\vec{u} \times \vec{B}_e + \vec{F}_s. \tag{16.3b}$$

The first equation expresses the *power balance* (see e.g. [22]). It follows from the second by scalar multiplication of the latter with $\vec{u}/u^0 = d\vec{z}/dt$.

If we choose the *initial conditions*

$$\vec{z}(0) = \vec{0}, \quad \dot{\vec{z}}(0) = \vec{u}(0) =: \vec{u}_0 = u_0 \vec{e}_1, \tag{16.4}$$

the sought solution of (16.3b),

$$\vec{z}(\tau) = x(\tau)\vec{e}_1 + y(\tau)\vec{e}_2, \tag{16.5}$$

has no 3-component.

For simplicity, we restrict the following calculation onto the *nonrelativistic case* $u_0 \ll 1$. Then we need not distinguish between time $t$ and proper time $\tau$. With (15.6), the *equation of motion* (16.3b) approximately reads at $\tau = 0$

$$m_0 \dot{\vec{u}}_0 \approx q \vec{u}_0 \times \vec{B}_e + m_1 \int_{-\infty}^{0} d\tau' \ddot{\vec{u}}(\tau') \exp\frac{-|\tau'|}{r_1}, \quad \dot{\vec{u}}_0 := \dot{\vec{u}}(0), \quad m_0 = m_\infty + m_1 . \tag{16.6}$$

The calculation takes the simplest form in *complex notation*:

$$z(\tau) = x(\tau) + i y(\tau), \quad u(\tau) := \dot{z}(\tau) = \dot{x} + i \dot{y}. \tag{16.7}$$

Then, the *initial conditions* (16.4) read

$$z(0) = 0, \quad \dot{z}(0) =: u_0 . \tag{16.8}$$

**Fig. 8**  We'll show that a positively charged particle spirals inward around the asymptotic position $z_\infty := z(\infty)$; see Fig. 8.

The pointers of the *complex Lorentz force* $F_e = i q u B_e$ and of the *proper acceleration* $\dot{u}$ lie in the $x, iy$-plane. Therefore, $F_e$ is a complex multiple $m$ of $\dot{u}$:

$$m \dot{u} = F_e = i q u B_e . \tag{16.9}$$

We call $m$ the *complex (effective) mass* of the particle. The *solution* of (16.9) is (for $\tau$-independent $m$)

$$u = u_0 e^{i\omega\tau} \quad \text{with} \quad \omega := \frac{q B_e}{m} =: \omega_r + i \omega_i = \frac{q B_e}{m_r + i m_i} . \tag{16.10}$$

For an inward spiraling particle, the *complex angular frequency* $\omega$ must have a positive imaginary part. That is, the imaginary part of $m$ must be negative:

$$\omega_i > 0 \iff m_i < 0 . \tag{16.11}$$

The initial (proper) acceleration is

$$a_0 := \dot{u}(0) = i \omega u_0 = \frac{i q u_0 B_e}{m} . \tag{16.12}$$

*Asymptotic cases* are [see (11.10a,b)]

$$|a_0| \ll <r>^{-1}: \quad \omega \approx \omega_0 := \frac{q B_e}{m_0} ; \tag{16.13a}$$

$$|a_0| \gg <r>^{-1}: \quad \omega \cong \omega_\infty := \frac{q B_e}{m_\infty} . \tag{16.13b}$$

In both borderline cases the spiral degenerates to a circle with radius $R_0 := u_0 / \omega_0$ and $R_\infty := u_0 / \omega_\infty$, respectively.

## 17.  Complex angular frequency and mass

To calculate the complex angular frequency $\omega$, we have to insert the solution (16.10) of the differential equation (16.9) into the *complex form of the equation of motion* (16.3b):

$$m_\infty \dot{u} = i q u B_e + F_s . \tag{17.1}$$

Then this equation has to be solved for the single unknown $\omega$ (or $m = q B_e / \omega$). The advantage of this procedure is that we *reduce the solution of the differential-integral equation (8.2) to the solution of the algebraic equation (17.1) for $\omega$*.

To simplify the calculation, let us restrict to small acceleration:

$$\frac{1}{c}|\dot{u}| = \left|\frac{quB_e}{mc}\right| \ll \langle r \rangle^{-1}. \qquad (17.2)$$

With $\langle r \rangle$ corresponding to (4.11) and (4.7), one gets for electrons and a magnetic field strength $B_e = 1\,T = 1\,Vs\,m^{-2}$ the restraint

$$|u| \ll 7 \times 10^9. \qquad (17.3)$$

If the inequality (17.2) is fulfilled, the *equation of motion* (16.6) is valid at $\tau = 0$. Its *complex form* reads

$$m_0 \dot{u}_0 \approx iqu_0 B_e + m_1 \int_{-\infty}^{0} d\tau'\, \ddot{u}(\tau)\exp\frac{-|\tau'|}{r_1}, \qquad \dot{u}_0 := \dot{u}(0). \qquad (17.4)$$

We now insert (16.10) in (17.4), calculate the integral and get (after some elementary transformations) the equation

$$(1-\mu_1)\Omega^2 - \left(1 + \frac{i\mu_1}{b}\right)\Omega + \frac{i\mu_1}{b} = 0, \qquad (17.5a)$$

with

$$\mu_1 := \frac{m_1}{m_0} = \frac{q^2}{8\pi r_1 m_0} \stackrel{(13.4)}{=} \frac{r_0}{r_1}, \qquad b := r_0 \omega_0 = \frac{r_0 q B_e}{m_0}. \qquad (17.5b)$$

$$\Omega := \frac{\omega}{\omega_0} = \frac{m_0}{m} =: \Omega_r + i\Omega_i. \qquad (17.5c)$$

For $\mu_1 \to 0$, the quadratic equation (17.5a) yields two limits: $\Omega \to 0$ and $\Omega \to 1$. Only the last is consistent with (17.5c). It gives $m := m_0/\Omega \to m_0$. The corresponding *solution* of (17.5a) reads for $\mu_1 < 0.5$

$$\Omega(\mu_1, b) = \frac{1}{2(1-\mu_1)}\left\{1 + \sqrt{\frac{w-\alpha}{2}} + i\left[\frac{\mu_1}{b} - \mathrm{sgn}\, b\sqrt{\frac{w+\alpha}{2}}\right]\right\}, \qquad (17.6a)$$

where

$$\alpha := \frac{\mu_1^2}{b^2} - 1 = \frac{1}{r_1^2 \omega_0^2} - 1 = \left(\frac{m_0}{r_1 q B_e}\right)^2 - 1, \quad \beta := 2\mu_1 - 1, \quad w := \sqrt{\alpha^2 + 4\mu_1^2 \beta^2/b^2}. \qquad (17.6b)$$

**Fig. 9** In Fig. 9 some polar diagrams of $\Omega(\mu_1, b)$ are drawn for $\mu_1 = const$ or $b = const$.

*Special cases* of (17.6a) are

1. $\omega_0 = \dfrac{qB_e}{m_0} = 0 \;\Rightarrow\; \omega = \omega_0 \Omega = 0,\; u \stackrel{(16.10)}{=} u_0.$ \hfill (17.7a)

2. $\mu_1 = \dfrac{m_1}{m_0} \ll 1 \;\Rightarrow\; \Omega := \dfrac{\omega}{\omega_0} := \dfrac{m_0}{m} = 1 + \left(1 + \dfrac{i}{b}\right)\mu_1 + O(\mu_1^2).$ \hfill (17.7b)

3. $b := \dfrac{q^3 B_e}{8\pi m_0^2} \ll 1 \;\Rightarrow\; \Omega = 1 + ib + \dfrac{1-2\mu_1}{\mu_1}b^2 + O(b^3).$ \hfill (17.7c)

4. $b \to \infty \;\Rightarrow\; \Omega \to \dfrac{1}{1-\mu_1} = \dfrac{m_0}{m_\infty} =: \dfrac{1}{\mu_\infty}.$ \hfill (17.7d)

The last limit is also correct if the inequality (17.2) is no more fulfilled [cf. (16.13b)]. Therefore, we suppose that (17.6a) is also a good approximation for arbitrary accelerations $\dot{u}$.

From the particle's speed $u = \dot{z}(\tau)$ (16.10), its *position* $z(\tau)$ immediately follows by (proper) time integration:

$$z(\tau) = \int_0^\tau d\tau' \dot{z}(\tau') = \frac{u_0}{i\omega}\left(e^{i\omega\tau} - 1\right), \quad \text{with} \quad \omega = \frac{qB_e}{m_0}\Omega. \tag{17.8}$$

The *asymptotic point* of this spiral is (see Fig. 8)

$$z_\infty := z(\infty) = \frac{iu_0}{\omega} = \frac{iu_0 m_0}{qB_e \Omega(\mu_1, b)}. \tag{17.9}$$

In the relativistic case ($u_0$ arbitrary), the trajectory (17.8) has the same form. But the complex angular frequency $\omega$ may depend on $u_0^2$.

That we had chosen $\tau = 0$ in (17.4) for the calculation of $\Omega$ did not mean a restriction of generality. The *generalization* of (17.4) for $\tau \neq 0$ is the equation of motion

$$m_0 \dot{u} \approx iquB_e + m_1 \int_{-\infty}^{\tau} d\tau' \ddot{u}(\tau') \exp\frac{-|\tau - \tau'|}{r_1} \quad \text{or} \tag{17.10}$$

$$m_\infty \dot{u} \approx iquB_e - \frac{m_1}{r_1} \int_{-\infty}^{\tau} d\tau' \dot{u}(\tau') \exp\frac{-|\tau - \tau'|}{r_1}. \tag{17.11}$$

Insertion of (16.10) into (17.10) or (17.11) leads to the same equation (17.5a) for $\Omega$.

The *differential-integral equations* (17.10) and (17.11) are linear homogeneous in $u$. Therefore, the initial speed $u_0$ drops out, so that $\Omega$ is independent of $u_0$.

Applying the differential operator

$$e^{-\tau/r_1} r_1 \frac{d}{d\tau} e^{\tau/r_1} \tag{17.12}$$

onto the differential-integral equation (17.10) transfers it into the *differential equation*

$$m_0 \dot{u} - iqB_e u = -r_1 \frac{d}{d\tau}\left(m_\infty \dot{u} - iqB_e u\right). \tag{17.13}$$

From it we can also derive the $\Omega$-determining equation (17.5a) by insertion of (16.10).

## 18  Forces, masses and powers

Let us *summarize the results* of the preceding sections. We had approximately solved the complex equation of motion (17.1) under the restraint (17.2). The approximate complex equations of motion (17.10) and (17.11) are satisfied by the solution $u = u_0 e^{i\omega\tau}$ (16.10) with the complex angular frequency $\omega = \omega_0 \Omega$, where $\Omega$ is given by (17.6a). The proper time derivative of the complex proper speed $u$ gives the complex proper acceleration

$$a = \dot{u} = i\omega u. \tag{18.1}$$

The equation of motion (17.1) may also be written in form of Newton's law $F_e = ma$ (16.9) with the *complex mass*

$$m \stackrel{(17.5c)}{=} m_0/\Omega = m_0 \omega_0/\omega = qB_e/\omega. \tag{18.2}$$

The *complex self-force* is

$$F_s := m_\infty a - F_e \stackrel{(17.11)}{=} -\frac{m_1}{r_1} \int_{-\infty}^{\tau} d\tau' \dot{u}(\tau') \exp\frac{-|\tau - \tau'|}{r_1}. \tag{18.3}$$

Reiterated integration by parts gives

$$F_s = -m_1(a - r_1 \dot{a} + r_1^2 \ddot{a} - \cdots) = -\frac{m_1}{1 + r_1 d/d\tau} a = -\frac{m_1}{1 + ir_1 \omega} a =: -m_s a \tag{18.4a}$$

with the *complex electrodynamic mass*
$$m_s = \frac{m_1}{1+ir_1\omega}, \qquad m_1 := m_0 - m_\infty . \tag{18.4b}$$
From this the *complex reactive force* immediately follows:
$$F_r := F_s + m_1 a =: m_r a , \tag{18.5}$$
with the *complex reactive mass*
$$m_r = m_1 - m_s = \frac{ir_1\omega m_1}{1 + ir_1\omega} . \tag{18.6}$$

The *complex mass* $m$ is connected with $m_s$:
$$m = m_\infty + m_s . \tag{18.7}$$
Inserting (18.2) and (18.4b), we again get eq. (17.5a) for the determination of the complex angular frequency $\omega = \omega_r + i\omega_i = \omega_0 \Omega$.

The *total energy* of the spiraling mass $m_0$ is
$$W_0(\tau) = m_0 \gamma := m_0 \left(1 + |u|^2\right)^{1/2}, \qquad |u| = u_0 \exp(-\omega_i \tau) , \tag{18.8}$$
with $\omega_i := \omega_0 \Omega_i = (qB_e / m_0)\Omega_i$ and $\Omega_i$ corresponding to (17.6a). This result is easier and physically more plausible than the approximation in [34, p. 235].

## 19. Invariance of the equation of motion (8.2) under any full Poincaré transformation

We postulate (in concordance with Tamm and Markov [36]) that *all fundamental equations of electrodynamics (in compact form) shall be invariant under any full Poincaré transformation*. That is, especially, they shall be invariant under any full (also nonorthochronous) Lorentz transformation. This implies the *invariance under a time inversion*. Let us interpret these *transformations* as *passive* [56]: The same physical process is described with respect to different reference frames of the Poincaré group. For the Maxwell equations this postulate is, as you know, fulfilled. It is also fulfilled for the equation of motion (9.12), valid for small magnitude $|\bar{a}|$ of acceleration.

The group of full Poincaré transformations (inhomogeneous Lorentz transformations) may be defined by the postulate: The 4-differential
$$d\bar{x} = dx^i \bar{e}_i = dx'^j \bar{e}'_j = d\bar{x}'$$
of the 4-position vector $\bar{x}$ be invariant under any transformation of the group.

A *time inversion* $T$ [11,28,57,58] is a special group element. It may be defined by the *transformation law for the coordinates* $x^i$ of the $T$-invariant 4-position vector
$$\bar{x} = x^i \bar{e}_i = x_i \bar{e}^i =: x'^i \bar{e}'_i = \bar{x}' : \tag{19.1}$$
$$T x^i =: x'^i = -\eta_{ij} x^j = -x_i , \tag{19.2}$$
with the *metric*
$$(\eta_{ij}) := (\bar{e}_i \cdot \bar{e}_j) = \text{diag}(1, -1, -1, -1) = (\bar{e}^i \cdot \bar{e}^j) =: (\eta^{ij}) .$$
From (19.1) and (19.2) follows the *transformation law for the base vectors*:
$$\bar{e}'_i = -\bar{e}^i , \quad \text{i.e.} \quad \bar{e}'_0 = -\bar{e}_0, \quad \bar{e}'_1 = \bar{e}_1, \quad \bar{e}'_2 = \bar{e}_2, \quad \bar{e}'_3 = \bar{e}_3 . \tag{19.3}$$
Contrarily follows (19.2) from (19.3) and (19.1), or (19.1) from (19.2) and (19.3).

The *passive time inversion*, used here, differs from a (more frequently used) active one in that *not only the time coordinate $x^0$ is inverted, but also the corresponding base vector $\bar{e}_0$*. All spacetime events $\bar{x}$ and worldlines $\bar{x}(\tau)$ remain invariant under the passive $T$-inversion. Natural processes are described with backward running clocks. A natural process is not inverted, but its timely description.

The time differential $dt$ and the proper time differential $d\tau := \sqrt{(d\bar{x}/dt)^2}\, dt$ have the same $T$-eigenvalue $-1$:

$$T\, dt = T\, dx^0 = -dx^0 = -dt, \qquad T\, d\tau = -d\tau.$$

The same holds for all odd proper time derivatives of $\bar{x}$:

$$T\bar{u}(\bar{x}) := T\frac{d\bar{x}}{d\tau} =: T\dot{\bar{x}} = \frac{d\bar{x}}{-d\tau} = -\bar{u}(\bar{x}) = -\dot{\bar{x}},$$

$$T\dddot{\bar{x}} = -\dddot{\bar{x}}, \ldots,$$

whereas even derivatives of $\bar{x}$ have the $T$-eigenvalue $+1$:

$$T\ddot{\bar{x}} = \ddot{\bar{x}}, \qquad T\frac{d^4}{d\tau^4}\bar{x} = \frac{d^4}{d\tau^4}\bar{x}, \quad \ldots.$$

The *4-current density* $\bar{j} := \rho_{el}\bar{u}$ is odd like the 4-velocity $\bar{u}$:

$$T\bar{j} = -\bar{j},$$

because the proper charge density $\rho_{el} = (\bar{j}^2)^{1/2}$ is $T$-even.

Application of the above postulate onto the wave equation

$$\partial_{\bar{x}}^2 \bar{A} = \bar{j}$$

yields that the *4-Potential* $\bar{A}$ must have the $T$-eigenvalue $-1$, like $\bar{j}$:

$$T\bar{A} = -\bar{A}.$$

The same is valid for the *electromagnetic field strength* $\bar{\bar{B}}$:

$$T\bar{\bar{B}} = T(-\partial_{\bar{x}}\bar{A} + \bar{A}\,\overleftarrow{\partial}_{\bar{x}}) = -\bar{\bar{B}}. \tag{19.4}$$

Hence, the *4-force density* $\bar{f}_\sigma := \rho_{el}\bar{u}\cdot\bar{\bar{B}}$ is even:

$$T\bar{f}_\sigma = \bar{f}_\sigma. \tag{19.5}$$

The integration over the $T$-invariant 3-plane $\sigma = \{\bar{x}\,|\,x^0 = 0\}$ does not change the $T$-eigenvalue $+1$ of $\bar{f}_\sigma$:

$$T\int_\sigma d\sigma\,\bar{f}_\sigma = T\int dq\,\bar{u}\cdot\bar{\bar{B}} = T\bar{F} = \bar{F}.$$

That is, the *4-force* $\bar{F} = \bar{F}_e + \bar{F}_s$ is *even*, too, like also the external 4-force $\bar{F}_e$ and the 4-self-force $\bar{F}_s$. [In (8.2), the following quantities are $T$-odd: $\bar{k}$, $i$, $d^4\bar{k}$. This results, e.g., from $\bar{p} = m\bar{u} = \hbar\bar{k} \triangleq -i\hbar\partial/\partial\bar{x}$]. Hence, the *equation of motion (8.2) is full Lorentz invariant*.

This does not hold for the Lorentz-Dirac equation [31,52,54]

$$m_0\dot{\bar{u}} - \bar{F}_e = \bar{F}_r, \tag{19.6a}$$

with

$$\bar{F}_r := \frac{q^2}{6\pi}\left(\ddot{\bar{u}} + \dot{\bar{u}}^2\,\bar{u}\right). \tag{19.6b}$$

The left side of (19.6a) is $T$-even, the right is odd. From the $T$-invariance postulate would follow $\bar{F}_r = \bar{0}$. For a homogeneous electric field this is a well-known (unphysical) result [45,54,60].

The opposite conclusion in [54] and [55] is caused by other definitions of the time-inverted fields and forces, and by the renunciation of the postulate that all forces must be $T$-even. Especially, eq. (9.20) in [55] is not compatible with our $T$-odd 4-potential (19.7).

The $T$-invariance postulate is not only fulfilled by our relativistic equation of motion (8.2), but also by its nonrelativistic approximation in [7] and [36] and by its approximation (16.6) for small acceleration.

That the 4-force density $\bar{f}_\sigma$ must be $T$-even is also provable in the following way: The equation of motion for a (charged) continuum of mechanical mass density $\mu_\infty$ reads

$$\mu_\infty \dot{\bar{u}} = \bar{f}_\sigma .$$

As the left side is $T$-even, the right must likewise be even. From this immediately follows, that $\bar{\bar{B}}$ must be odd:

$$\bar{f}_\sigma - T\bar{f}_\sigma = (1-T)\rho_{el}\bar{u}\cdot\bar{\bar{B}} = \rho_{el}\bar{u}\cdot(\bar{\bar{B}} + T\bar{\bar{B}}) = \bar{0} \quad \Rightarrow \quad T\bar{\bar{B}} = -\bar{\bar{B}} .$$

The *Liénard-Wiechert field* $\bar{\bar{B}}$ [56,57] of an accelerated point charge $q$ is a well-known example: Considering that

$$T R := T\bar{u}\cdot\bar{r} = -\bar{u}\cdot\bar{r} = -R , \quad \text{with} \quad \bar{r} := \bar{x} - \bar{z}, \quad \bar{r}^2 = 0 ,$$

one gets the $T$-odd *retarded 4-potential* [54]

$$\bar{A}(\bar{x}) = \frac{q\bar{u}}{4\pi|R|} , \tag{19.7}$$

from which the odd *retarded electromagnetic field strength*

$$\bar{\bar{B}}(\bar{x}) := -\partial_{\bar{x}}\bar{A} + \bar{A}\overleftarrow{\partial}_{\bar{x}}$$

$$= \frac{q}{4\pi R|R|}\left[\bar{a}\bar{r} - \bar{r}\bar{a} - \frac{\bar{a}\cdot\bar{r}-1}{R}(\bar{u}\bar{r} - \bar{r}\bar{u})\right] =: q\bar{\bar{b}} \tag{19.8}$$

follows. The 4-velocity $\bar{u}$ and the 4-acceleration $\bar{a} := \dot{\bar{u}}$ are to be taken at the retarded 4-point $\bar{z}$. This event is full Lorentz invariant just as the 4-point $\bar{x}$ of observation.

In a proper (orthochronous) Lorentz frame, the retarded 4-point $\bar{z}$ is defined as the first penetration point of the particle's worldline $\bar{z}(\tau)$ through the light cone $(\bar{x}-\bar{z})^2 = 0$ (with its vertex at $\bar{x}$). For it holds $R := \bar{u}\cdot(\bar{x}-\bar{z}) > 0$.

The retarded 4-potential $\bar{A}$ and the retarded electromagnetic field strength $\bar{\bar{B}}$ depend both explicitly and implicitly on $\bar{x}$, the latter because of $\bar{u} = \bar{u}(\bar{z})$, $\bar{a} = \bar{a}(\bar{z})$ and $(\bar{x}-\bar{z})^2 = 0$. (In [10] and [34] this fact is ignored.)

The self-field $\bar{\bar{B}}_s$ can be represented as superposition of elementary waves (19.8), emitted by retarded charge elements $dq'$:

$$\bar{\bar{B}}_s = \int d\bar{\bar{B}}_s = \int dq' \bar{\bar{b}} .$$

Thus, the mean electromagnetic self-field strength is

$$\langle\bar{\bar{B}}_s\rangle := \frac{1}{q}\int_\sigma dq\int dq'\,\bar{\bar{b}} . \tag{19.9}$$

Therefore, $\langle\bar{\bar{B}}_s\rangle$ is odd and the self-force $\bar{F}_s := q\bar{u}\cdot\langle\bar{\bar{B}}_s\rangle$ even. The former also follows from (7.6).

To guarantee the full Lorentz invariance of electrodynamics, we postulate:

*Each (compact) field tensor $F$ (of rank 0, 1, 2, ... , i.e. scalar, 4-vector, 4-dyadic, ...) shall be an eigentensor of the eigenvalue equation*

$$L F = \pm F \tag{19.10}$$

*with the eigenvalue +1 or −1*, where $L$ is any full (homogeneous) Lorentz transformation.

This demand is equivalent to the postulate: *Each field tensor $F$ shall commute or anticommute with the Lorentz operator $L$*:

$$L F \mp F L =: 0. \tag{19.11}$$

In our paper, we preferred the *compact representation* of physical quantities against their coordinate representation, otherwise more frequently used, because the former is *simpler* and more lucid. For example, the electromagnetic field strength in compact notation, $\bar{\bar{B}}$, is $T$-odd. On the contrary, in contravariant coordinate notation, $(B^{i,j})$ (7.5) is no eigenmatrix of the time inversion operator $T$.

The *equation of motion* (8.2) and its equivalent (8.5) are not only invariant under any full Lorentz transformation, but also under any 4-translation in spacetime. Hence, they are *invariant under any full Poincaré transformation*, too.

The full Poincaré invariance also holds, as you know, for the Maxwell equations. The claim that this invariance shall be conserved, if one adds to the electric charges hypothetical *magnetic charges* (see, e.g., [60]) with $T$-even density $\rho_m$, yields $\rho_m = 0$. The contrary assumption that a (time independent) magnetic charge $q_m$ is $T$-odd is physically implausible.

## 20.    Properties of the solutions of the equation of motion

From the equation of motion (8.2) or its integrated form (8.5), the following properties of the solutions can be inferred:

1.    *A preacceleration does not occur*.

That is: If the external 4-force $\bar{F}_e = \bar{0}$ for $\tau \leq 0$ and is $\neq \bar{0}$ for $\tau > 0$, then this also holds for the 4-acceleration $\bar{a}$ and the 4-self-force $\bar{F}_s$ (9.2), which only depends on the past (see App. D for the special case $f := qE_e/m = 0$). Hence,

$$\bar{F}_e = q\bar{u} \cdot \bar{\bar{B}}_e = \bar{0} \quad \text{for} \quad \tau \leq 0 \;\Rightarrow\; \bar{a} = \dot{\bar{u}} = \bar{0} \quad \text{for} \quad \tau \leq 0. \tag{20.1}$$

2.    *A postacceleration does occur*.

That is: If the external 4-force $\bar{F}_e \neq \bar{0}$ for $\tau \leq 0$ and is $= \bar{0}$ for $\tau > 0$, then the latter does not hold for the 4-acceleration $\bar{a}$. This follows from

$$\bar{a}(+0) \stackrel{(8.1)}{=} \frac{1}{m_\infty} \bar{F}_s(0) \neq \bar{0}, \tag{20.2}$$

because the self-force is a continuous quantity of process, depending on the past.

## 21.    Discussion

As advantages of the equation of motion (8.2) or its equivalent (8.5) against other ones (cf., e.g., [3,35,39,42,43,46, 47,54,56,61,62,63]) we want to name (see also App. C):

1. For the derivation of these differential-integral equations only Maxwell's continuum electrodynamics has been used. The Maxwell equations were not changed (in contrast, e.g., to [17]). Expansions of nonanalytic functions in powers of a small parameter (for example the pseudo radius $r_1$) were not applied.

2. The electromagnetic 4-self-field strength $\bar{\bar{B}}_s$, used for the calculation of the 4-self-force $\bar{F}_s$, is the retarded solution of Maxwell's equations belonging to the 4-current density $\bar{j}$. With that the causality principle is preserved.

3. It was dispensed with the divergent, mathematically and physically unacceptable mass renormalization usual for point charges. That is, Dirac's postulate (see preamble) is satisfied.

4. It was also dispensed with additional ad hoc boundary conditions at infinity, like $\bar{a} := \bar{0}$ for $t = \pm\infty$, to exclude runaway solutions. The latter, just as noncausal preacceleration [21], do not occur. Instead of that, a postacceleration occurs [see Sect. 20].

5. The magnitudes $|\bar{F}_e|$ and $|\bar{a}|$ of the external 4-force and the 4-acceleration need not be very small. Of course, quantum mechanical effects, like pair creation, are not described by our classical theory. The contrary assumption in [18] is contestable.

6. For the homogeneous electric field holds: If one of the 4-vectors $\bar{a}$, $\bar{F}_e$, $\bar{F}_s$, $\bar{F}_r$ vanishes, this also holds for the other three [cf. (13.2) and Fig. 5]. This inference from our solution of the equation of motion (8.2) [approximated by (9.11)] is physically more plausible than that in case of the Lorentz-Dirac equation (G6a,b), where $\bar{F}_r := m_0 \bar{a} - \bar{F}_e = \bar{0}$ also holds for $|\bar{F}_e| = const \neq 0$. The notorious discrepancy, that a uniformly accelerated charged particle radiates and simultaneously the reaction force vanishes, disappears [37, 45]. Instead of this, our equation (8.2) yields the physically more plausible result that $\bar{F}_r \neq \bar{0}$ for the hyperbolic motion.

7. The differential-integral equation (8.2) is causal. That is, the solution $\bar{z}(\tau)$ only depends on the past. The equation (8.2) simplifies for small $|\bar{a}|$ and $|\bar{F}_e|$ to $m_0 \bar{a} \approx \bar{F}_e$, for large $|\bar{a}|$ and $|\bar{F}_e|$ to $m_\infty \bar{a} \simeq \bar{F}_e$. Our conclusions are consistent with those of Markov [36], who nonrelativistically calculated the response of an extended charge to an oscillating external force.

8. The asymptotic equations of motion are especially fulfilled by our solutions for the motion of the particle in a homogeneous electric or magnetic field.

9. Like Maxwell's equations and the asymptotic equations $m_0 \bar{a} \approx \bar{F}_e$ and $m_\infty \bar{a} \simeq \bar{F}_e$, the equations of motion (8.2) and (8.5) are invariant under any *full* Poincaré transformation (not only under any orthochronous one, see Sect. 19). This does not hold for the Lorentz-Dirac equation and its generalizations in [17] and [18]. Nor it holds for the Caldirola-Yaghjian equation [55,66] and the equations, proposed by Eliezer, Mo and Papas, Herrera, and Bonnor ([43], p. 211).

10. The equation of motion (8.2), written in compact notation, is not only forminvariant (= covariant) under full Poincaré transformations, but also invariant, because both sides are.

11. For our examples with
$$\bar{\bar{B}}_e = E_e(\bar{e}_0\bar{e}_1 - \bar{e}_0\bar{e}_1) = \overline{\overline{const}} \quad \text{or} \quad \bar{\bar{B}}_e = B_e(\bar{e}_1\bar{e}_2 - \bar{e}_2\bar{e}_1) = \overline{\overline{const}}$$
amazingly simple and physically reasonable solutions result. It is to conjecture that this also holds for other external electromagnetic field strengths $\bar{\bar{B}}_e(\bar{x})$.

12. For rectilinear motions follows from $\bar{u} \cdot \bar{F}_e = 0$ and $\bar{u} \cdot \bar{a} = 0$ that the equation of motion (8.2) may be written in the form $m\bar{a} = \bar{F}_e$. The effective inertial mass $m = m_\infty - \bar{F}_s/\bar{a}$ is in general a functional of $\bar{z}(\tau' < \tau)$ [cf. (8.2)]. In the case of a homogeneous electric field treated above, $m$ is a time-independent nonanalytic function of $|\bar{a}|$.

13. If it would be possible to measure the dependence $m(|\bar{a}|) = \bar{F}_e/\bar{a}$, one could derive from this function the density function $\rho(r')$ (4.4).

14. If the particle moves in a plane, it is advantageous to go over to complex quantities $z(\tau)$, $a = \ddot{z}(\tau)$, $F_e$, $m$, .... Then, the complex equation of motion $ma = F_e$ is valid additionally to (8.2). With this method, the spiraling particle in a homogeneous magnetic field $B_e$ was treated. In this case, $m(B_e)$ (18.2) is time-independent.

The relativistic equation of motion $m_0 \dot{\bar{u}} \approx q \bar{\bar{u}} \cdot \bar{\bar{B}}_e$ holds for a small, extended charge $q$ of inertial mass $m_0 =: m_\infty + m_1$, moving in a weak external electromagnetic field $\bar{\bar{B}}_e$. The 4-self-force $\bar{F}_s \approx -m_1 \dot{\bar{u}}$ may be combined with the mechanical inertial 4-force $-m_\infty \dot{\bar{u}}$ to the total inertial 4-force $-m_0 \bar{a}$. For a very strong electromagnetic field $\bar{\bar{B}}_e$, in the above weak field equation, $m_0$ has to be replaced by $m_\infty$.

The accelerated heavy mass part $m_h$ of $m_\infty = m_h + m_g$ is a source of a gravitational wave effecting a gravitational 4-self-force $-m_g \dot{\bar{u}}$, which is a tiny part of the mechanical inertial 4-force $-m_\infty \dot{\bar{u}}$.

If in the local rest frame $R''$ acts, additionally to the external electric field strength $\vec{E}''_e$, a small external gravitational field strength $\vec{g}''_e$, we can consider this by the ansatz $\vec{F}''_e = q\vec{E}''_e + m_\infty \vec{g}''_e$.

That the extended charge does not explode, although enormous electromagnetic repulsive forces act between the charge elements, is explainable by the assumption that nonelectromagnetic forces compensate these repulsive forces. With this (Poincaré) model conception, the notorious factor 4/3 in (9.10) is explainable, too [39,59].

The product $r_1 \bar{F}_s(r_1)$ of the particle's quasi-radius with the 4-self-force is nonanalytic at $r_1 = 0$ [cf. (12.8)]. This is an essential cause for the fact that the equation of motion (8.2) does not go over into the Lorentz-Dirac equation [47,54] for $r_1|\bar{a}| \ll 1$ (see App. F).

Any equation of motion should for $|\bar{a}| \to 0$ go over into $m_0 \bar{a} \approx \bar{F}_e$ (9.12). This claim is in general not fulfilled by other equations of motion, e.g. the Lorentz-Dirac equation, if $|\dot{\bar{a}}| \neq 0$.

The consideration of the reaction force in quantum theory is possible by substitution of the external 4-force $\bar{F}_e$ by the effective force $\bar{F}_{e,eff}(\bar{F}_e)$ [cf. (14.7)].

## Acknowledgment

I thank my son Peter Stöckel for technical support.

# Appendices

## A. Electrostatic self-energy of the charged particle

To prove the claim after (9.3), we calculate the electrostatic self-energy of the charged particle in spherical coordinates:

$$m_1 = \frac{q^2}{8\pi} \int d^3\vec{r}' \rho' \int_0^\infty dr\, r^2 \int_0^\pi d\vartheta\, 2\pi \sin\vartheta\, \rho \left(r^2 + r'^2 - 2rr'\cos\vartheta\right)^{-1/2}$$

$$= \frac{q^2}{8\pi} \int dr'\, 4\pi r'^2 \rho' \int_0^\infty dr\, 4\pi r^2 \rho \left[\frac{\Theta(r'-r)}{r'} + \frac{\Theta(r-r')}{r}\right]$$

$$= \frac{1}{8\pi} \iint \frac{dq'\, dq}{r_>} , \tag{A1}$$

where

$$\vartheta = \angle(\vec{r},\vec{r}'), \quad \rho = \rho(r), \quad \rho' := \rho(r'), \quad \Theta(x) = \text{unit step function} = \int_{-\infty}^x dy\, \delta(y),$$

$$dq = q\rho\, 4\pi r^2\, dr, \quad dq' = q\rho'\, 4\pi r'^2 dr', \quad r_> := \text{Max}(r, r').$$

The integral over $dr$ may be transformed with the help of (4.4):

$$\int_0^\infty dr\ldots = \frac{4\pi}{r'} \int_0^{r'} dr\, r^2 \rho + 4\pi \int_{r'}^\infty dr\, r\rho$$

$$= \frac{1}{i\pi} \int_{-\infty}^\infty d\kappa\, \kappa\, \tilde{\rho}(\kappa) \left(\frac{1}{r'}\int_0^{r'} dr\, r\, e^{i\kappa r} + \int_{r'}^\infty dr\, e^{i\kappa r}\right) = \frac{1}{\pi r'} \int_{-\infty}^\infty d\kappa\, \tilde{\rho}\, \frac{\sin\kappa r'}{\kappa}.$$

Insertion in (A1) yields

$$m_1 = \frac{q^2}{8\pi} \int_0^\infty dr'\, 4\pi r'^2\, \frac{1}{4\pi^2 r'} \int_{-\infty}^\infty d\kappa'\, \kappa'\, \tilde{\rho}'\, \sin\kappa' r'\, \frac{1}{\pi r'} \int_{-\infty}^\infty d\kappa\, \frac{\tilde{\rho}}{\kappa} \sin\kappa r'$$

$$= \frac{q^2}{8\pi^3} \int d\kappa' \int d\kappa\, \frac{\kappa'}{\kappa}\, \tilde{\rho}'\tilde{\rho} \int_0^\infty dr\, \sin\kappa' r\, \sin\kappa r$$

$$= \frac{q^2}{8\pi^3} \iint d\kappa'\, d\kappa\, \frac{\kappa'}{\kappa}\, \tilde{\rho}'\tilde{\rho}\, \frac{\pi}{2} \left[\delta(\kappa'-\kappa) - \delta(\kappa'+\kappa)\right]$$

$$= \frac{q^2}{8\pi^2} \int_{-\infty}^\infty d\kappa\, \tilde{\rho}^2, \tag{A2}$$

in agreement with (9.10).

The calculated interaction energy between the charge elements equals the electrostatic self-field energy (see, e.g., [28,34]).

## B. Proof that the field strength balance (12.5) holds for an arbitrary reference point of the particle's worldline

For the derivation of the field strength balance (12.5), we had supposed as a reference point $\bar{z} = \bar{0}$. We now want to show that the same balance results for an arbitrary reference point $\bar{z}_1$ of the particle's worldline (12.1), too.

For the position vector $\bar{z}$ of an arbitrary point on the particle's worldline holds [see (D6)]

$$\bar{z} = z^0 \bar{e}_0 + z^1 \bar{e}_1 = t\bar{e}_0 + x\bar{e}_1 = \frac{\sinh f\tau}{f} \bar{e}_0 + \frac{\cosh f\tau - 1}{f} \bar{e}_1 . \tag{B1}$$

At this, $\bar{e}_0$ and $\bar{e}_1$ are the orthonormal base vectors of the lab frame, pointing into the positive directions of the $t$ and $x$ axis, respectively, with the origin at $\bar{z} = \bar{0}$ (Fig. 10). The proper 2-velocity at a fixed point $\bar{z}_1 := \bar{z}(\tau_1)$ is

$$\dot{\bar{z}}(\tau_1) =: \bar{U} = \bar{e}_0 U^0 + \bar{e}_1 U^1 = \bar{e}_0 \cosh f\tau_1 + \bar{e}_1 \sinh f\tau_1 . \tag{B2}$$

**Fig. 10** Additionally to the lab frame, we introduce the local rest frame with its origin at $\bar{z}_1$. Its base vectors $\bar{e}'_0 = \bar{U}$ and $\bar{e}'_1 = \bar{e}_0 U^1 + \bar{e}_1 U^0$ point into the positive directions of the $t'$ and $x'$ coordinate axes. The inversion of this *Lorentz transformation of the base vectors* reads

$$\bar{e}_0 = U^0 \bar{e}'_0 - U^1 \bar{e}'_1 ,$$

$$\bar{e}_1 = -U^1 \bar{e}'_0 + U^0 \bar{e}'_1 .$$

With that, one gets

$$\bar{z}' := \bar{z} - \bar{z}_1 = (t - t_1)\bar{e}_0 + (x - x_1)\bar{e}_1$$

$$= f^{-1}[(\sinh f\tau - \sinh f\tau_1)(\bar{e}'_0 \cosh f\tau_1 - \bar{e}'_1 \sinh f\tau_1)$$

$$+ (\cosh f\tau - \cosh f\tau_1)(-\bar{e}'_0 \sinh f_1\tau_1 + \bar{e}'_1 \cosh f\tau_1)]$$

$$= \frac{\sinh f\tau'}{f} \bar{e}'_0 + \frac{\cosh f\tau' - 1}{f} \bar{e}'_1 , \quad \text{where} \quad \tau' := \tau - \tau_1 . \tag{B3}$$

This 2-position vector has in all local rest frames the same form: It is forminvariant under any proper-time translation $\tau \to \tau - \tau_1$. Equation (B1) is the special case for $\bar{z}_1 = \bar{0}$. This specialization means, therefore, no restriction for the general validity of (12.5).

Taking into account the Poincaré invariance of the equation of motion (8.2), the above result implies the conclusion: *The solution (B1) fulfils this equation not only at $\bar{z} = \bar{0}$, but also at an arbitrary trajectory point $\bar{z} = \bar{z}_1$.*

## C. Comparison with Nodvik's work

In a fundamental work, Nodvik [40] derived an equation of motion for an extended charge in an electromagnetic field. He did not restrict himself to a nonrotating charge with not too large an acceleration. The complexity of his differential-integral equation may be the reason that (to the author's knowledge) no nontrivial solutions have been found up to now.

To simplify the problem, we restricted ourselves from the beginning to a nonrotating charge and to $a \ll <r>^{-1}$ (3.6). A further simplification was the passing over to Fourier transformed quantities, like usual in quantumelectrodynamics.

Independently of Nodvik´s derivation, we found the equation of motion (8.2), valid under the above-mentioned conditions. For this equation, nontrivial (approximate) solutions were found. The asymptotic forms for small and large acceleration of the differential-integral equation and its solutions were discussed.

If we had not approximated the functional determinant $D = 1 - \vec{r} \cdot \vec{a}$ (3.3) by 1 and $\tilde{\rho}'$ by $(3/4)\tilde{\rho}$ (9.11), our corresponding generalization of (8.2) should accord with Nodvik´s equation of motion, specialized on a nonrotating charge.

### D. Example for the iterative solution of the equation of motion (8.5)

For a homogeneous electric field $\vec{E}_e = E_e \vec{e}_1$, we get from (8.1) and (11.8a)

$$\dot{\bar{u}} = \frac{q}{m_\infty} \bar{u} \cdot <\bar{\bar{B}}> = \frac{q}{m} \bar{u} \cdot \bar{\bar{B}}_e \quad \text{or} \tag{D1}$$

$$\frac{q}{m_\infty} <\bar{\bar{B}}> = \frac{q}{m} \bar{\bar{B}}_e =: f(\bar{e}_0 \bar{e}_1 - \bar{e}_1 \bar{e}_0) =: f \bar{\bar{d}} =: \bar{\bar{f}}, \quad f := \frac{qE_e}{m}. \tag{D2}$$

In the following we show that in this case the *iteration procedure converges against the unique solution (D4)*.

Assuming the *initial conditions*

$$\bar{z}(0) = \bar{0}, \quad \dot{\bar{z}}(0) = \bar{e}_0 \tag{D3}$$

and beginning with the zeroth approximation

$$\bar{z}^{(0)} = \bar{0},$$

the differential-integral equation (8.5) successively yields (with $\bar{e}_0 \cdot \bar{e}_0 = -\bar{e}_1 \cdot \bar{e}_1 = 1$, $\bar{e}_0 \cdot \bar{e}_1 = 0$)

$$\bar{z}^{(1)} = \tau \bar{e}_0,$$

$$\bar{z}^{(2)} = \tau \bar{e}_0 + \frac{f\tau^2}{2} \bar{e}_1,$$

$$\bar{z}^{(3)} = \left(\tau + \frac{1}{6} f^2 \tau^3\right) \bar{e}_0 + \frac{f\tau^2}{2} \bar{e}_1,$$

$$\bar{z}^{(4)} = \left(\tau + \frac{1}{3!} f^2 \tau^3\right) \bar{e}_0 + \left(\frac{1}{2!} f \tau^2 + \frac{1}{4!} f^3 \tau^4\right) \bar{e}_1.$$

The *n*th approximation $\bar{z}^{(n)}$ converges for $n \to \infty$ against the well-known hyperbola

$$\bar{z}(\tau, f) = \lim_{n \to \infty} \bar{z}^{(n)} = \frac{\sinh f\tau}{f} \bar{e}_0 + \frac{\cosh f\tau - 1}{f} \bar{e}_1 = t \bar{e}_0 + x \bar{e}_1. \tag{D4}$$

From this follows (12.1):

$$x = \frac{1}{f}(\cosh f\tau - 1) = \frac{1}{f}\left(\sqrt{1 + \sinh^2 f\tau} - 1\right) = \frac{1}{f}\left(\sqrt{1 + f^2 t^2} - 1\right). \tag{D5}$$

The *solution* (D4) *of (8.5)* may also be written in the form (see [51])

$$\bar{z}(\tau, f) = \int_0^\tau d\tau' \, \dot{\bar{z}}(\tau', f) = \bar{e}_0 \cdot \int_0^\tau d\tau' \exp(\tau' \bar{\bar{f}}) = \bar{e}_0 \cdot \frac{\exp(\tau \bar{\bar{f}}) - \bar{\bar{1}}}{\bar{\bar{f}}}$$

$$:= \bar{e}_0 \cdot \left[ \tau \bar{\bar{1}} + \frac{f\tau^2}{2!} (\bar{e}_0 \bar{e}_1 - \bar{e}_1 \bar{e}_0) + \frac{f^2 \tau^3}{3!} (\bar{e}_0 \bar{e}_1 - \bar{e}_1 \bar{e}_0) \cdot (\bar{e}_0 \bar{e}_1 - \bar{e}_1 \bar{e}_0) + \cdots \right]$$

$$= \bar{e}_0 \frac{\sinh f\tau}{f} + \bar{e}_1 \frac{\cosh f\tau - 1}{f}. \tag{D6}$$

The mean field strength dyadic $\bar{\bar{f}}$ (D2) is the *generator* [22] *of the Lorentz transformation dyadic* (= $\tau$-translation)

$$\bar{\bar{L}}(\tau \bar{\bar{f}}) = \exp(\tau \bar{\bar{f}}) = \exp(\tau f \bar{\bar{d}}) := \bar{\bar{1}} + \tau f \bar{\bar{d}} + \frac{1}{2} \tau^2 f^2 \bar{\bar{d}}^2 + \cdots, \tag{D7}$$

mapping the initial 4-speed $\dot{\bar{z}}(0, f) = \bar{e}_0$ onto the speed at proper time $\tau$:

$$\dot{\bar{z}}(\tau, f) = \dot{\bar{z}}(0, f) \cdot \bar{\bar{L}}(\tau \bar{\bar{f}}) = \bar{e}_0 \cdot \exp(\tau \bar{\bar{f}}) = \bar{e}_0 \cosh f\tau + \bar{e}_1 \sinh f\tau. \tag{D8}$$

At this we have used the identities

$$\bar{\bar{d}}^{2n+1} \equiv \bar{\bar{d}} := \bar{e}_0 \bar{e}_1 - \bar{e}_1 \bar{e}_0, \qquad n = 0, 1, 2, \ldots;$$

$$\bar{\bar{d}}^{2n+2} \equiv \bar{\bar{d}}^2 = \bar{e}_0 \bar{e}_0 - \bar{e}_1 \bar{e}_1 =: \bar{\bar{1}}, \tag{D9}$$

$$\bar{e}_0 \cdot \bar{\bar{d}}^{2n} \equiv \bar{e}_0, \qquad \bar{e}_0 \cdot \bar{\bar{d}}^{2n+1} \equiv \bar{e}_1.$$

$\bar{\bar{1}}$ is the unit dyadic in the two-dimensional 0, 1 Minkowski space.

The field strength $f := qE_e/m$ may be determined as a function of $f_\infty := qE_e/m_\infty$ like in Sect.12.

## E. Calculation of $\Delta(\delta)$ (12.8) with generalized factorial functions

The function $\Delta(\delta)$ (12.8) may be calculated, e.g., as follows.

Insertion of the power series (10.15) for $A_\parallel(\alpha)$ into (12.8) yields

$$\Delta(\delta) = \sum_{n=0,1,\ldots}^\infty c_{2n} f_{2n}(\delta), \tag{E1}$$

with

$$c_{2n} := 6 \frac{1+n}{(3+2n)!}, \qquad \Delta(\infty) = \sum_0^\infty c_{2n} f_{2n}(\infty) = \sum_0^\infty c_{2n} (2n)! = 1.5 \tag{E2}$$

and for $\delta \geq 0$:

$$f_n(\delta) := \int_0^\infty dt \, t^n \exp\left(-\sqrt{t^2 + 2t/\delta}\right) = \frac{2^n}{\delta^{n+1}} \int_0^\infty dx \, \sinh x \, \sinh^{2n} \frac{x}{2} \exp\left(-\frac{\sinh x}{\delta}\right), \tag{E3}$$

$$f_n(-\delta) := -f_n(\delta).$$

These *generalized factorial functions* increase with $\delta$ from $f_n(0) = 0$ to $f_n(\infty) = n!$. Their *asymptotic behavior* is

$$f_n(\delta) \approx \frac{(2n)!}{2^n}\delta^{1+n} \quad \text{for} \quad \delta \ll 1,$$
$$\simeq n!\,(1 - \delta^{-1} + \cdots) \quad \text{for} \quad \delta \gg 1. \tag{E4}$$

$f_n(\delta)$ resembles the incomplete factorial function $(n,\delta)! := \int_0^\delta dt\, t^n e^{-t}$ [27].

## F. The Lorentz-Dirac equation as an inadmissible approximation of the equation of motion (8.2)

The *Lorentz-Dirac equation* can be derived *as an approximation of (8.2)*. To show this, we start from eq. (9.2) for the self-force. With the approximation
$$\tilde{\rho}' \approx \tilde{\rho} \tag{F1}$$
instead of (9.4b), one gets, analogously to the (nonrelativistic) derivation of (9.10), instead of (9.8)
$$\bar{F}_s \approx -\frac{q^2}{6\pi^2}\int_{-\infty}^0 dt'\,\vec{u}' \int_{-\infty}^\infty d\kappa\, \kappa^2 \exp(i\kappa t')\,\tilde{\rho}^2. \tag{F2}$$
Inserting
$$\vec{u}' \approx \bar{a}\,t' + \dot{\bar{a}}\frac{t'^2}{2} \tag{F3}$$
gives with (A2) the *nonrelativistic self-force* at the origin $\bar{z} = \bar{0}$ of a local rest frame, where $\bar{u} = \bar{e}_0$:
$$\bar{F}_s \approx -\frac{4}{3}m_1\bar{a} + \frac{q^2}{6\pi}\dot{\bar{a}}, \qquad \bar{u}\cdot\bar{F}_s = 0. \tag{F4}$$
For $\bar{z} \ne \bar{0}$ and $\bar{u} = \dot{\bar{z}} \ne \bar{e}_0$, the *relativistic generalization* of this formula obviously reads
$$\bar{F}_s \approx -\frac{4}{3}m_1\bar{a} + \frac{q^2}{6\pi}\dot{\bar{a}}_\sigma \quad \text{with} \quad \dot{\bar{a}}_\sigma := \bar{\bar{P}}_\sigma\cdot\dot{\bar{a}} := (\bar{\bar{1}} - \bar{u}\,\bar{u})\cdot\dot{\bar{a}}, \quad \bar{u}\cdot\bar{F}_s = 0. \tag{F5}$$
The *projection dyadic*
$$\bar{\bar{P}}_\sigma := \bar{\bar{1}} - \bar{u}\,\bar{u} \tag{F6}$$
projects a 4-vector, here $\dot{\bar{a}}$, onto the 3-plane $\sigma \perp \bar{u}$ [see (2.9) and Fig. 1]. The approximate 4-self-force $\bar{F}_s$ (F5) yields the *Lorentz-Dirac equation* (19.6a,b),
$$\left(m_\infty + \frac{4}{3}m_1\right)\bar{a} = \bar{F}_e + \frac{q^2}{6\pi}\bar{\bar{P}}_\sigma\cdot\dot{\bar{a}}. \tag{F7}$$

Unfortunately, this "derivation" of the Lorentz-Dirac equation as an approximation of the equation of motion (8.2) is mathematically and physically inadmissible, and this because of the following reasons:

(1)  The product $<r>\bar{F}_s$ is in general a nonanalytic function of the radius $<r>$ [cf. (12.8)]. Therefore, $\bar{F}_s$ is not developable into a Laurent expansion in $<r>$. The approximation (F5) may be considered as a truncated (divergent asymptotic) Laurent series with $m_1 \propto <r^{-1}>$ [cf. (9.14)].

(2) The limit $\lim_{<r> \to 0} m_1 = \infty$. That means: The transition to a point charge is to exclude. The so-called mass renormalization with the ad hoc assumption $m_\infty \to -\infty$, so that the "observable mass" $m_\infty + 4m_1/3 = -\infty + \infty :< \infty$, is mathematically and physically unsound and contradicts Dirac's postulate (see preamble).

(3) The Lorentz-Dirac equation contains the relativistically incorrect "electrodynamic mass" $4m_1/3$, caused by the incorrect approximation (F1).

(4) The equation of motion (8.2) yields [cf. (9.10)]: Decreasing acceleration implies decreasing reaction force,
$$\overline{a} \to \overline{0} \quad \Rightarrow \quad \overline{F}_r := m_1 \overline{a} + \overline{F}_s := m_0 \overline{a} - \overline{F}_e \to \overline{0} \,. \tag{F8}$$
This physically necessary property of (8.2) is destroyed at the transition to the Lorentz-Dirac equation.

(5) The equation (8.2) also implies the inversion
$$\overline{F}_r \to \overline{0} \quad \Rightarrow \quad \overline{a} \to \overline{0} \tag{F9}$$
[cf. (13.2) and Fig. 5]. This physically senseful connection is destroyed, too. A well-known example is the hyperbolic motion, where in the Lorentz-Dirac equation $\overline{F}_r = \overline{0}$, although $\overline{a} \neq \overline{0}$, and although electromagnetic radiation is emitted.

Consequently, the Lorentz-Dirac equation (F5) cannot be considered as an admissible approximation of (8.2). Similar objections can be rendered to other derivations of the Lorentz-Dirac equation, for example to Dirac's [12]. His self-force is not the Lorentz force acted upon the particle's charge by the retarded electromagnetic self-field at proper time $\tau$.

## References

1. *M. Abraham*, Prinzipien der Dynamik des Elektrons, Ann. Physik 10 (1903) 105.
2. *M. Abramowitz* and *I.A. Stegun* (eds.), Handbook of mathematical functions, Dover Publications, New York, 1989.
3. *W. Appel, M. K.-H. Kiessling,* Mass and spin renormalization in Lorentz electrodynamics, arXiv: math-ph/0009003 v2   8 Oct 2000.
4. *A.O. Barut*, Brief history and recent developments in electron theory and quantum-electrodynamics, in *D. Hestenes and A. Weingartshofer* (eds.), The electron, new theory and experiment, fundamental theories of physics, Dordrecht et al., 1991.
5. *W.W. Batygin* and *I.N. Toptygin*, Aufgabensammlung zur Elektrodynamik, Deutscher Verlag der Wissenschaften, Berlin, 1965.
6. *R.G. Beil*, The extended classical charged particle, Found. Phys. 19 (1989) 319.
7. *A. Belousov*, J. Exper. Theor. Phys. 9 (1939) 658.
8. *D. Bohm* and *M. Weinstein*, Phys. Rev. 74 (1948) 1789.
9. *M. Born*, Ann. Physik 30 (1909) 1.
10. *A.E. Chubykalo* and *S.J. Vlaev*, Necessity of simultaneous co-existence of instantaneous and retarded interactions, Intern. J. Mod. Phys. A 14 (1999) 3789-3798.
11. *A. Das*, The special theory of relativity, Springer, Berlin et al., 1993.

12. *P.A.M. Dirac*, Classical theory of radiating electrons, Proc. Roy. Soc. London A <u>167</u> (1938) 148.
13. *P.A.M. Dirac*, The relativistic electron wave equation, Europhysics news <u>8</u> (10) (1977) 1.
14. *C.J. Eliezer*, Proc. Roy. Soc. London A <u>194</u> (1948) 543-555.
15. *T. Erber*, The classical theories of radiation reaction, Fortschr. Phys. <u>9</u> (1961) 343-392.
16. *E. Fermi,* Über einen Widerspruch zwischen der elektrodynamischen und der relativistischen Theorie der elektromagnetischen Masse, Physik. Zeitschrift <u>23</u>(1922)340
17. *J. Frenkel* and *R.B. Santes*, The self-force of a charged particle in classical electrodynamics with a cutoff, Int. J. Mod. Phys. B <u>13</u> (1999) 315-324.
18. *V.L. Ginzburg*, Applications of electrodynamics in theoretical physics and astrophysics, Gordon and Breach, New York et al., 1989.
19. *I.S. Gradstein* and *I.M. Ryshik*, Produkt- und Integraltafeln, Deutsch, Thun, Frankfurt am Main, 1981.
20. *W.T. Grandy, Jr.*, Introduction to electrodynamics and radiation, New York, London, 1970.
21. *W.T. Grandy, Jr.*, Relativistic quantum mechanics of leptons and fields, Dordrecht, Boston, London, 1991.
22. *W. Greiner* and *J. Rafelski*, Theoretische Physik, Band 3A, Spezielle Relativitätstheorie, Deutsch, Thun, Frankfurt am Main, 1984.
23. *H. Grotch, E. Kazes, F. Rohrlich, D.H. Sharp*, Internal retardation, Acta Phys. Austr. <u>54</u> (1982) 31-38.
24. *S.F. Gull*, Charged particles at potential steps, in *D. Hestenes* and *A. Weingartshofer* (eds.), The electron, new theory and experiment, fundamental theories of physics, Dordrecht et al., 1991.
25. *A. Gupta* and *T. Padmanabhan*, Radiation of a charged particle and radiation reaction reexamined, Phys. Rev. D <u>57</u> (1998) 7241-7250.
26. *G. Herglotz*, Ann. Physik <u>31</u> (1910) 393.
27. *V. Hnizdo*, American J. Phys. <u>66</u> (1998) 414-418.
28. *J.D. Jackson*, Klassische Elektrodynamik, Wiley, Berlin, New York, 1983.
29. *Jahnke* and *F. Emde*, Tafeln höherer Funktionen, Teubner, Leipzig, 1948.
30. *D.J. Kaup*, Classical motion of an extended charged particle, Phys. Rev. <u>152</u> (1966) 1130.
31. *K.-J. Kim*, The equation of motion of an electron: A debate in classical and quantum physics, Nucl. Instr. Meth. Phys. Res. A <u>429</u> (1999) 1- 8.
32. *H. Kolbenstvedt*, Electromagnetic self-mass of the classical electron. ..., Phys. Lett. A <u>234</u> (1997) 319-321.
33. *W. Kuhn, H. Stöckel* and *H. Glaßl*, Mathematische Hilfsmittel der Physik, Barth, Heidelberg, Leipzig, 1995.
34. *L.D. Landau* and *E.M. Lifschitz*, Klassische Feldtheorie, Akademie-Verlag, Berlin, 1971.
35. *H.A. Lorentz*, The theory of electrons, Teubner, Leipzig, 1909.
36. *M. Markov*, On the back action of the electromagnetic field on a moving electron, J. Phys. <u>10</u> (1946) 159.
37. *K.D. McDonald*, Limits on the applicability of classical electromagnetic fields as inferred from the radiation reaction., www.hep.princeton.edu/~mcdonald/accel/, (1998) 1-15.
38. *P.M. Morse* and *H. Feshbach*, Methods of theoretical physics, Mc Graw-Hill, New York et al., 1953.
39. *M. Nakajima*, Physical interpretation of the electromagnetic mass, Helv. Phys. Acta <u>71</u> (1998) 392-425.

40. *J.S. Nodvik*, A covariant formulation of classical electrodynamics for charges of finite extension, Annals Phys. 28 (1964) 225-319.
41. *E. Noether*, Ann. Physik 31 (1910) 919.
42. *K.H. Panofsky* and *M. Phillips*, Classical electricity and magnetism, Addison-Wesley, Reading et al., 1962.
43. *S. Parrott*, Relativistic electrodynamics and differential geometry, Springer, New York, 1987.
44. *S. Parrott*, Unphysical and physical (?) solutions of the Lorentz-Dirac equation, Found. Phys. 23 (1993) 1093-1119.
45. *S. Parrott*, Radiation from a charge accelerated for all time, Gen. Rel. Grav. 29 (1997) 1463-1472.
46. *S. Parrott*, Radiation from a uniformly accelerated charge and the equivalence principle, gr-qc/9303025 v6 (1998).
47. *P. Pearle*, Classical electron models, in *D. Teplitz* (ed.), Electromagnetism, Plenum, New York, 1983.
48. *L. de la Peña*, *J.L. Jimenez* and *R. Montmayor*, The classical motion of an extended charged particle revisited, Nuovo Cim. B 69 (1982) 71-88.
49. *G.N. Plass*, Classical electrodynamic equations of motion with radiation reaction, Rev. Mod. Phys. 33 (1961) 37.
50. *H. Poincaré*, Sur la dynamique de l'electron, C.R. Acad. Sci. Paris 140 (1905) 1504.
51. *A. Prieto,* On the electrodynamic group: The relativistic radiation reaction force, J. Math. Phys. 39 (1998) 1478.
52. *J.A.E. Roa-Neri* and *J.L. Jimenez*, A systematic approach to the Lorentz-Dirac equation, Nuovo Cim. B 111 (1996) 1051-1057.
53. *F. Rohrlich*, Self-energy and stability of the classical electron, Am. J. Phys. 28(1960)639.
54. *F. Rohrlich*, Classical charged particles, Addison-Wesley, Reading, Massachusetts, 1965.
55. *F. Rohrlich*, The arrow of time in the equations of motion, Found. Phys. 28 (1998) 1045-1056.
56. *H. Römer* and *M. Forger*, Elementare Feldtheorie, Verlag Chemie, Weinheim, 1993.
57. *G. Scharf,* From electrostatics to optics, Springer, Berlin et al., 1994.
58. *M. Schottenloher,* Geometrie und Symmetrie in der Physik, Braunschweig, 1995.
59. *J. Schwinger*, Electromagnetic mass revisited, Found. Phys. 13 (1983) 373.
60. *J. Schwinger, L.L. De Raad, K. Milton* and *W.-y. Tsai*, Classical electrodynamics, Addison-Wesley, Reading, Massachusetts, 1998.
61. *A.A. Sokolov* and *I.M. Ternov*, Synchrotron radiation, Deutscher Verlag der Wissenschaften, Berlin, 1968.
62. *H. Spohn,* Dynamics of charged particles and their radiation field, Cambridge University Press, 2004.
63. *H. Stöckel*, Bewegung einer Punktladung in elektromagnetischen Feldern unter Berücksichtigung der Strahlungsdämpfung, Fortschr. Phys. 24 (1976) 417-475.
64. *J.L. Synge*, Relativity: The general theory, North-Holland Publ. Comp., Amsterdam, 1964.
65. *J.L. Synge*, Relativity: The special theory, North-Holland Publ. Comp., Amsterdam, 1965.
66. *A.D. Yaghjian*, Relativistic dynamics of a charged sphere, Lecture notes in physics, Springer, Berlin et al., 1992.

## Legends

Fig. 1. Parallel worldlines of the particle centre $\bar{z}(\tau)$ and a particle-fixed point $\bar{x}(\tau)$. For simplicity, we have omitted the bars, indicating 4-vectors.

Fig. 2. Radial density $\rho_{1r} := \dfrac{dq}{q}\dfrac{r_1}{dr} = 4\pi r^2 r_1 \rho_1 = \dfrac{2}{\pi}\dfrac{r}{r_1} K_1\left(\dfrac{r}{r_1}\right)$ as a function of the reduced radius $r/r_1$.

Fig. 3. The contour $C_r$ is a semicircle in the upper complex $\omega$ half plane with arbitrarily large radius $R$.

Fig. 4. Hyperbolic worldlines (12.1), belonging to the external field strengths $f_0 := F_e/m_0$, $f = F_e/m$ and $f_\infty = F_e/m_\infty$.

Fig. 5. The connections (12.5) and (13.1) between the field strengths $f = a = F_e/m$, $f_\infty = F_e/m_\infty$, $f_s = F_s/m_\infty$, $f_0 = F_e/m_0$ and $f_r = F_r/m_0$: $f = f_\infty + f_s = f_0 + f_r$. Example: $m_0/m_\infty = 1.5$.

Fig. 6. $f_0(f)$ with $\mu_1 := m_1/m_0 = r_0/r_1$ as a parameter.

Fig. 7. Relative effective mass $\mu = m/m_0$ (14.2) as a function of $\delta_0 := r_0 f$ with the relative electrostatic mass $\mu_1 := m_1/m_0$ as a parameter.

Fig. 8. Spiral $z(\tau)$ (17.8) of $q > 0$ in a homogeneous external magnetic field $\vec{B}_e$.

Fig. 9. The complex relative frequency $\Omega(\mu_1, b) = \Omega_r + i\Omega_i$ (17.6a) for $\mu_1 = 0.05, 0.1, ..., 0.5$ (17.5b) and $b = 0.5, 1, ..., 3$, respectively.

Fig. 10. The $t, x$ and $t', x'$ coordinate axes span the local rest frames with the origins at $\bar{z} = \bar{0}$ and $\bar{z}_1$.

**Table 1.** The functions $\Delta(\delta)$ (12.8), $\mu_r(\delta) := 1 - \Delta/\delta$ (12.12) and $\tilde{\mu}_r(\delta)$ (12.14)

| $\delta$ | 0 | 0.1 | 0.2 | 0.5 | 1 | 5 | 10 | 100 | $\infty$ |
|---|---|---|---|---|---|---|---|---|---|
| $\Delta$ | 0 | 0.099999 | 0.1995 | 0.4624 | 0.7418 | 1.2577 | 1.3667 | 1.4719 | 1.5 |
| $\mu_r$ | 0 | 0.000010 | 0.00236 | 0.0752 | 0.2582 | 0.7485 | 0.8633 | 0.98528 | 1 |
| $\tilde{\mu}_r$ | 0 | 0.000014 | 0.00237 | 0.0740 | 0.2582 | 0.7511 | 0.8650 | 0.98527 | 1 |

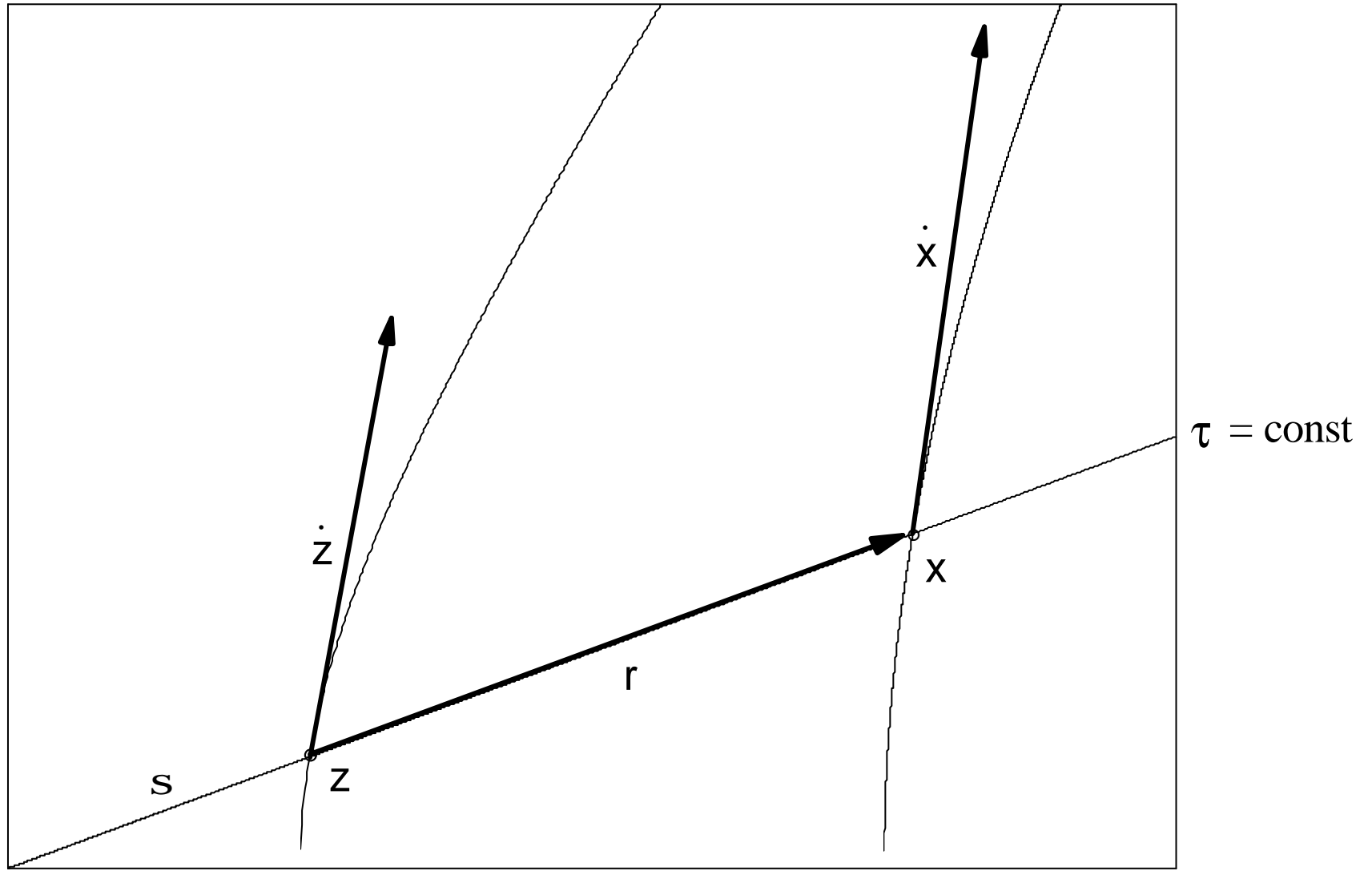

H. Stöckel: Fig. 1

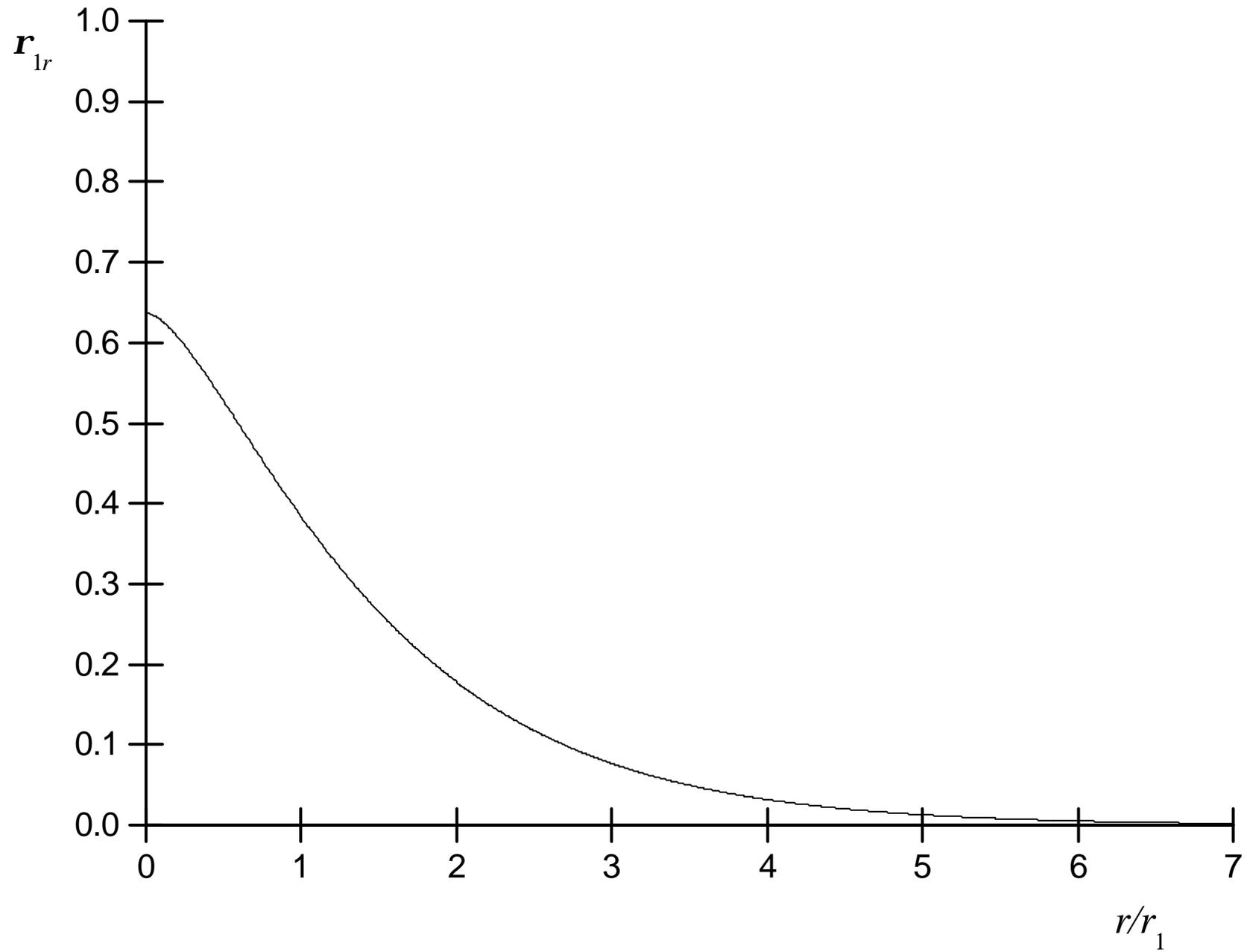

H. Stöckel: Fig. 2

H. Stöckel: Fig. 3

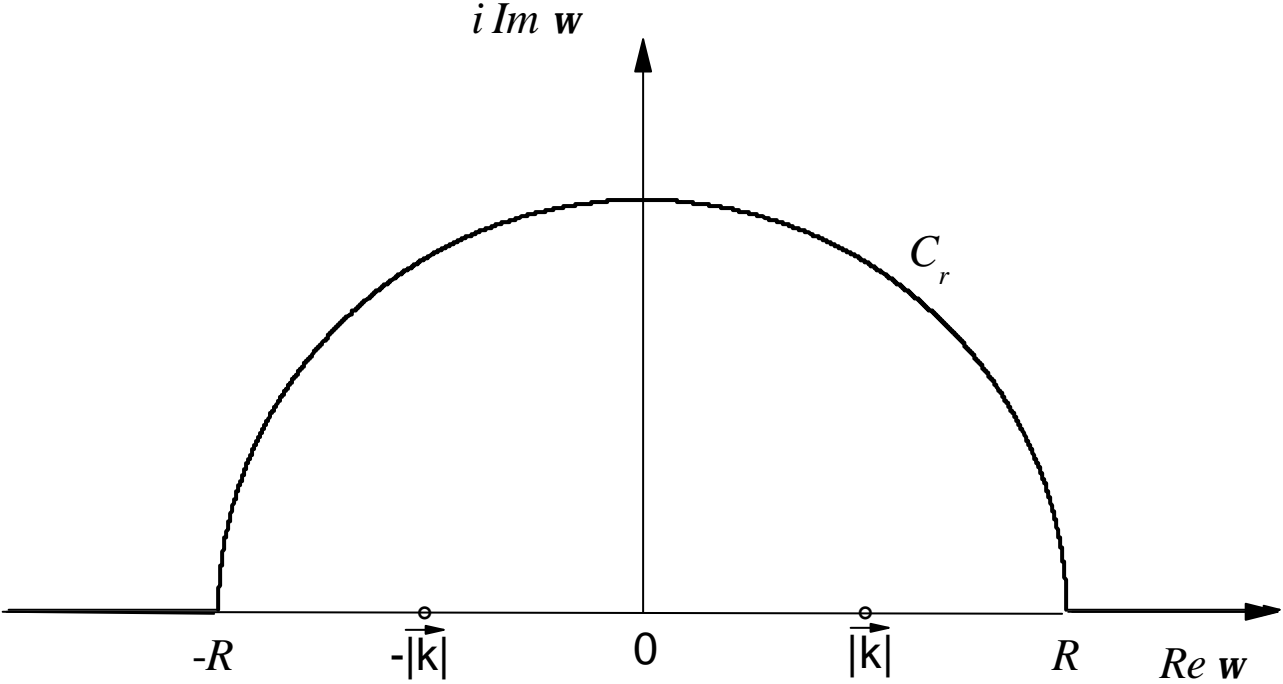

H. Stöckel: Fig. 4

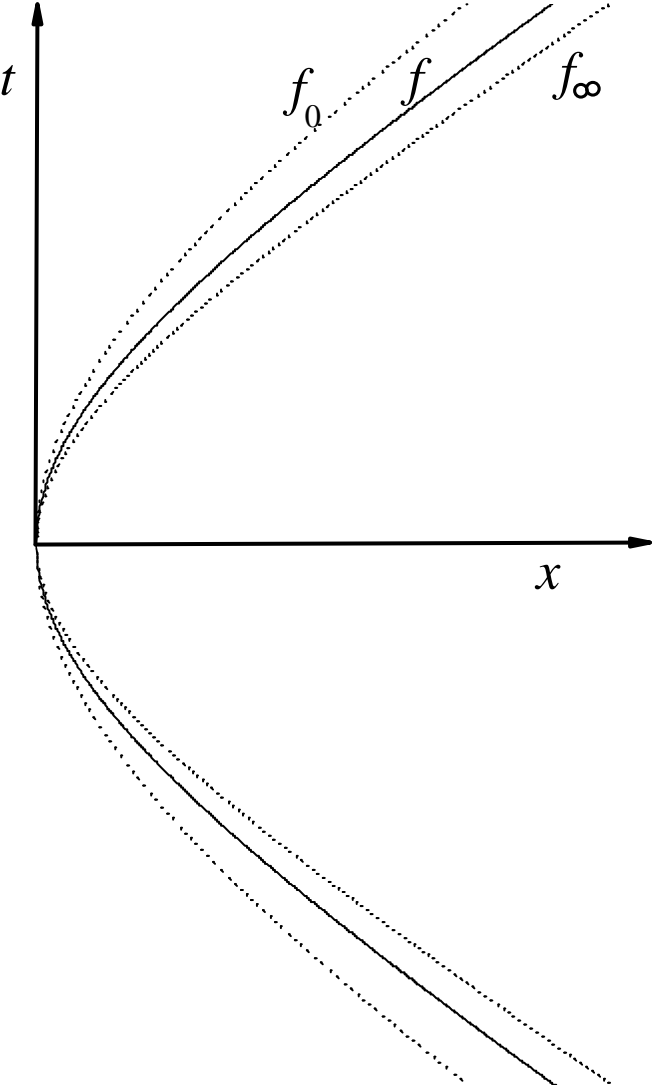

H. Stöckel: Fig. 5

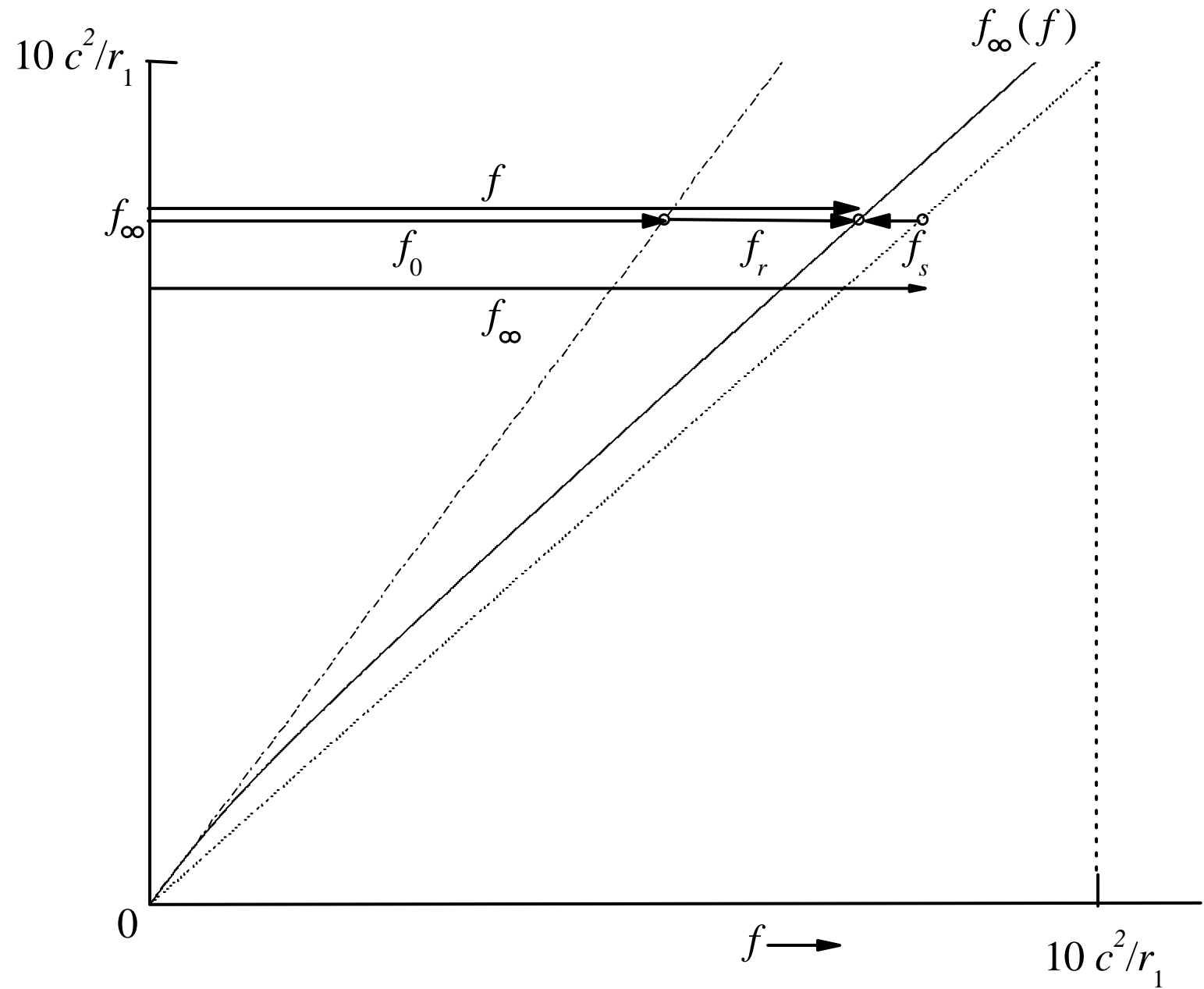

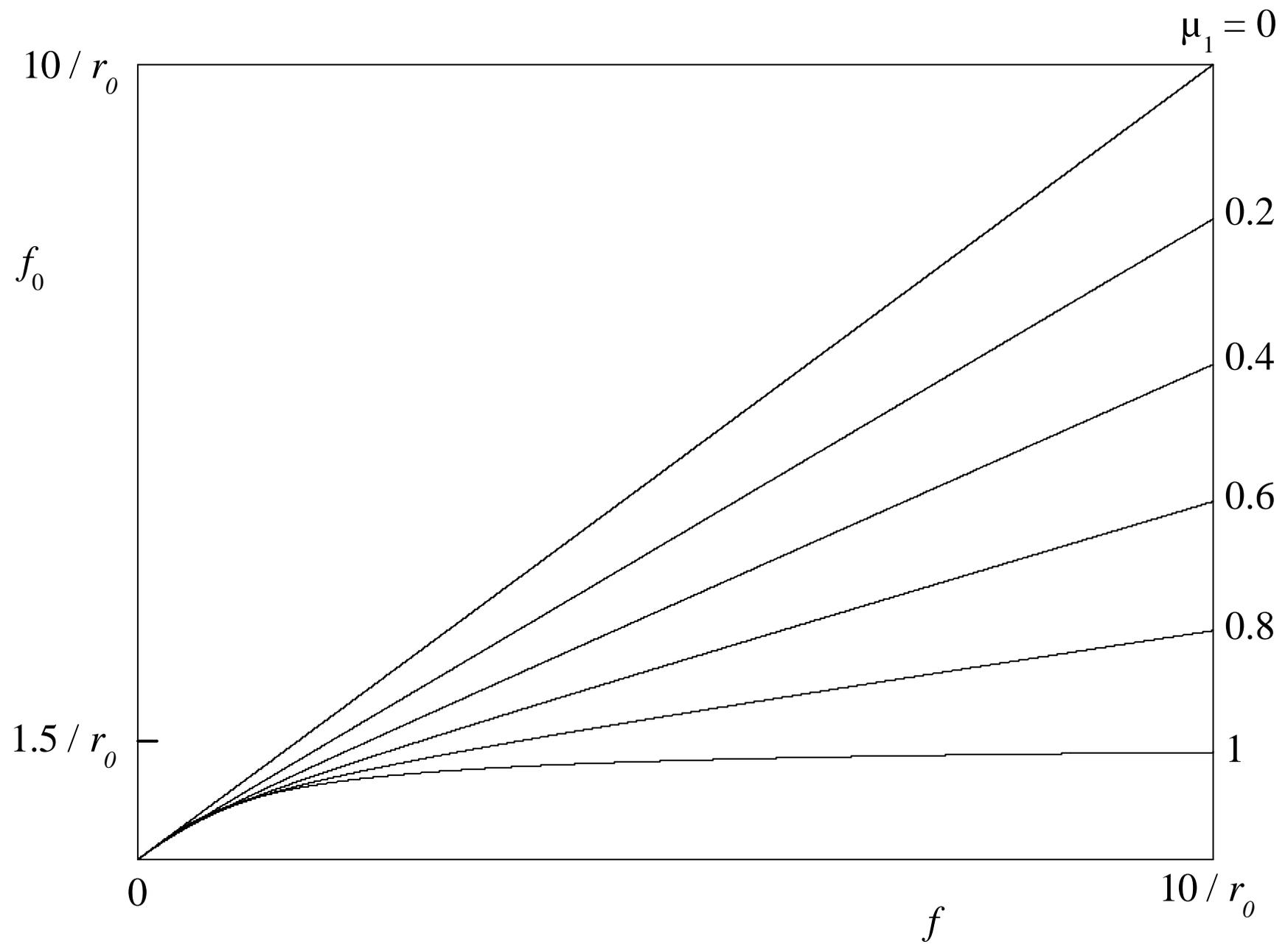
H. Stöckel: Fig. 6

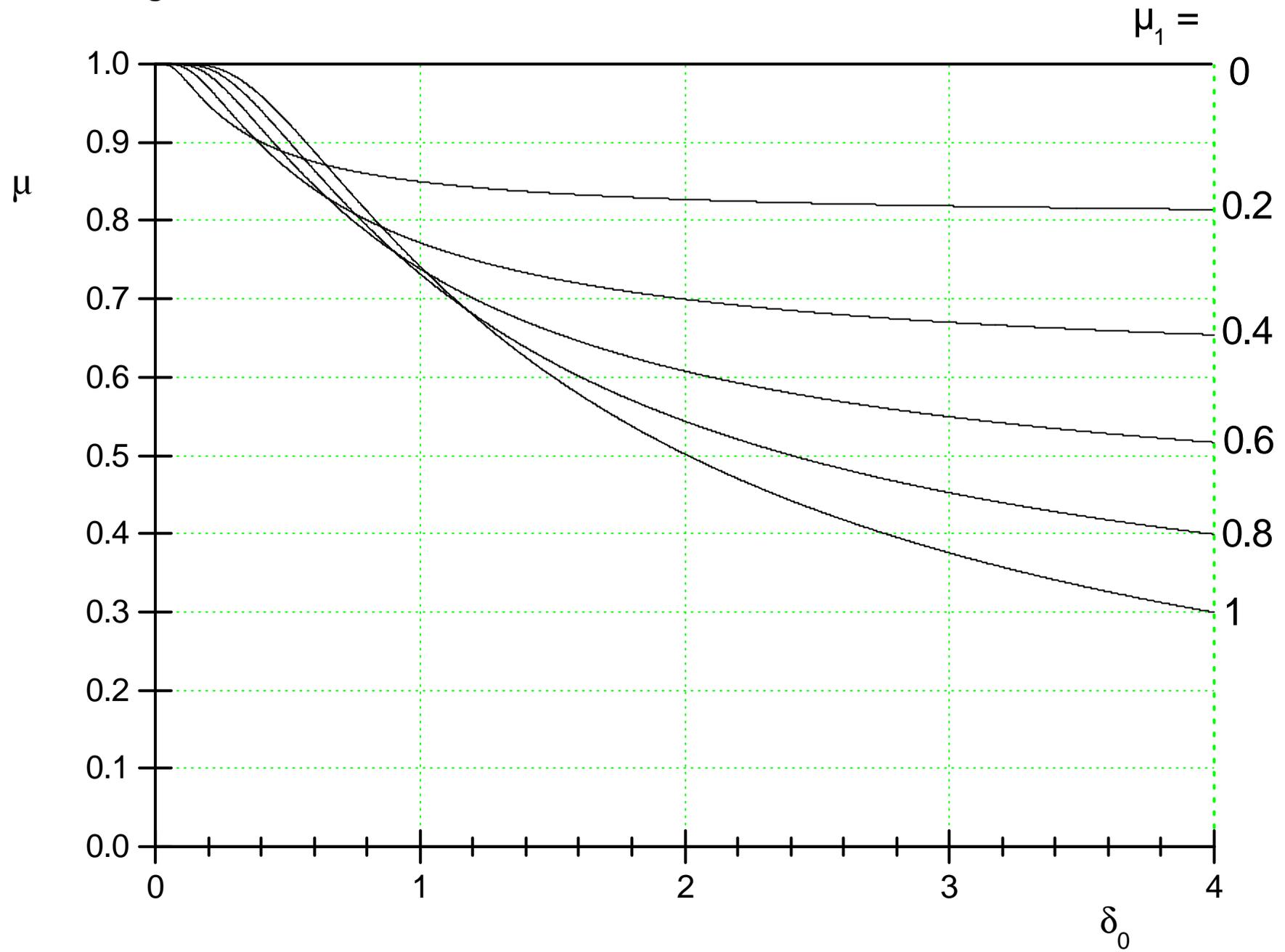

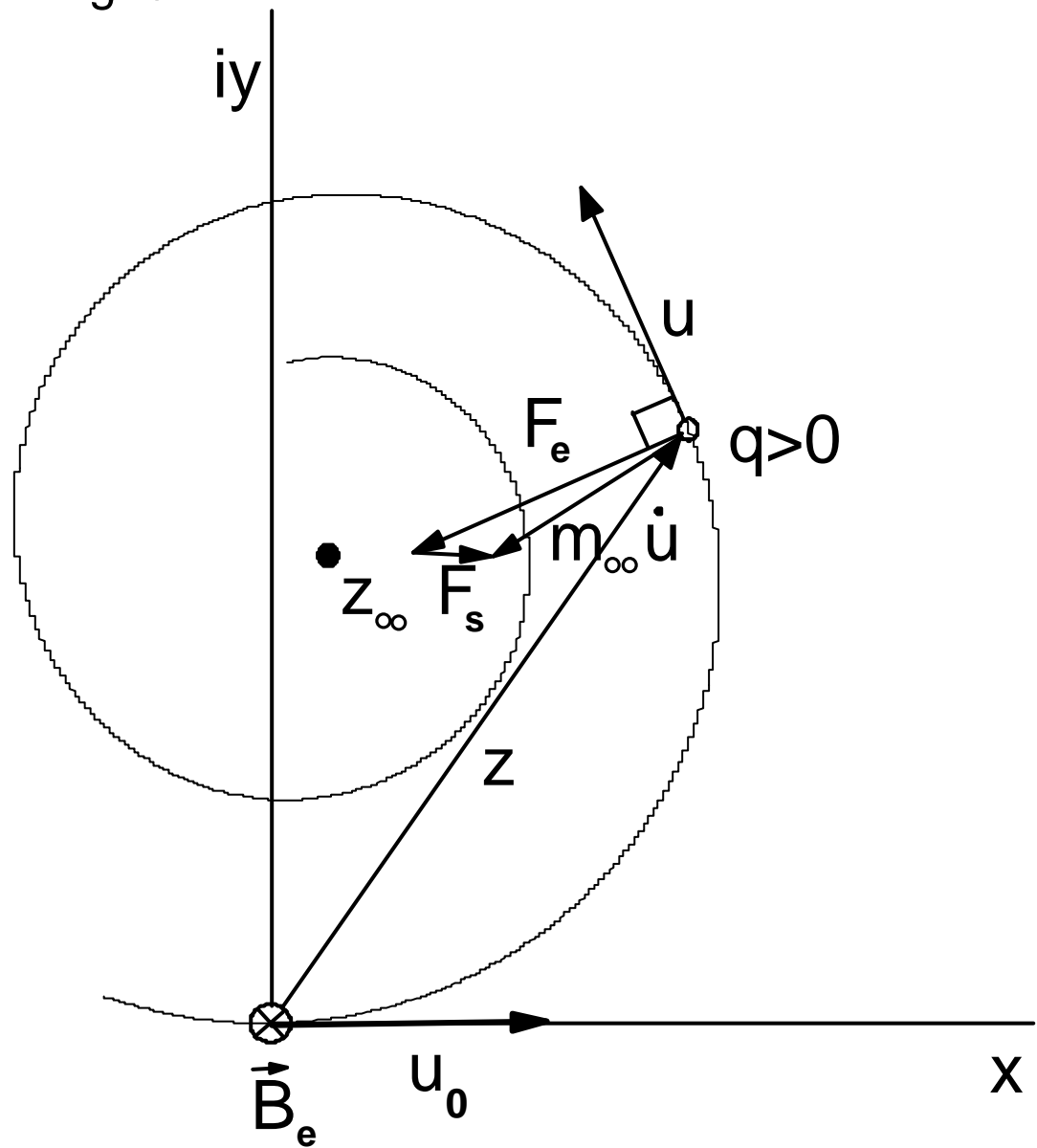

H. Stöckel: Fig. 8

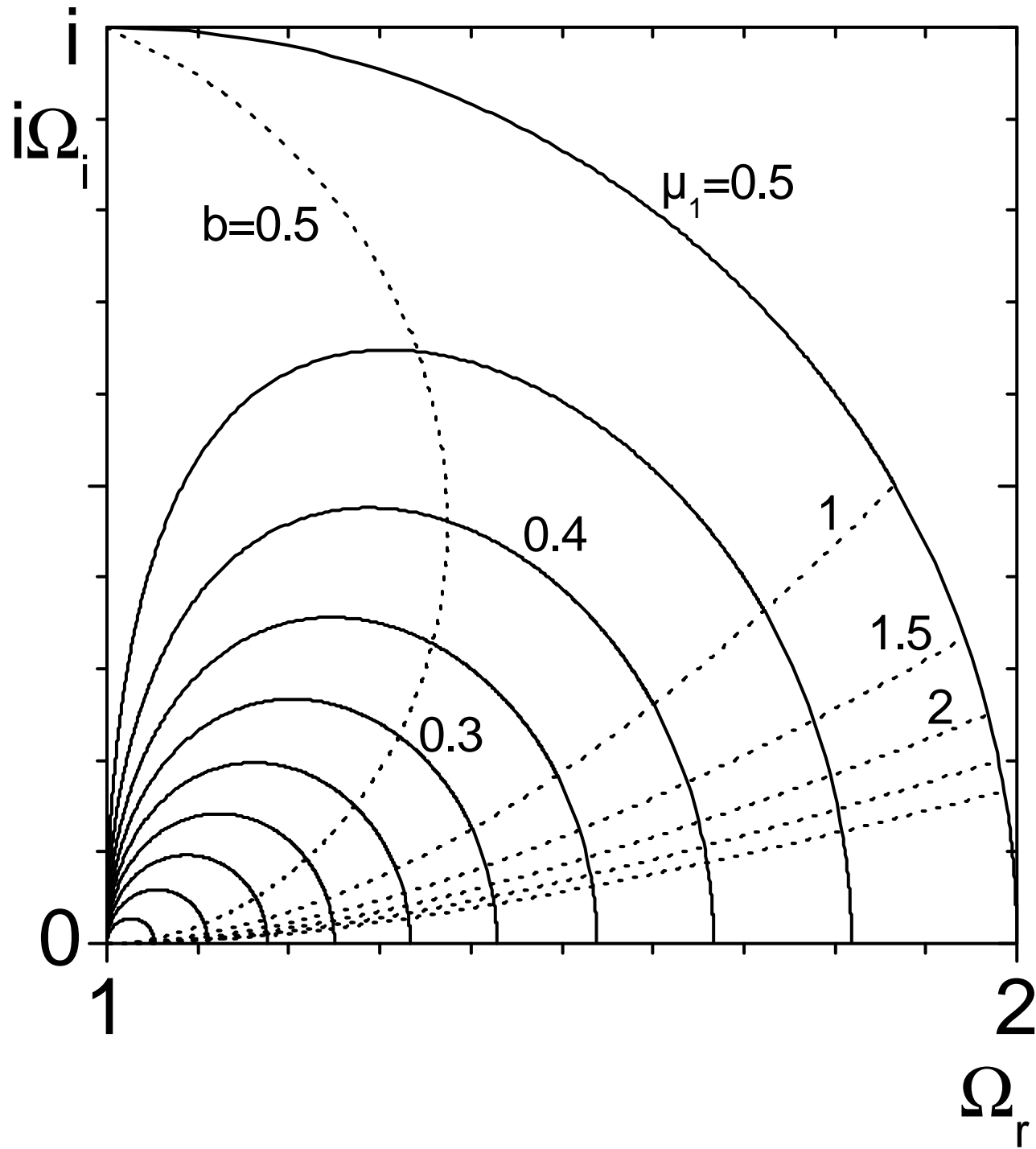

H. Stöckel: Fig. 9

H. Stöckel: Fig. 10

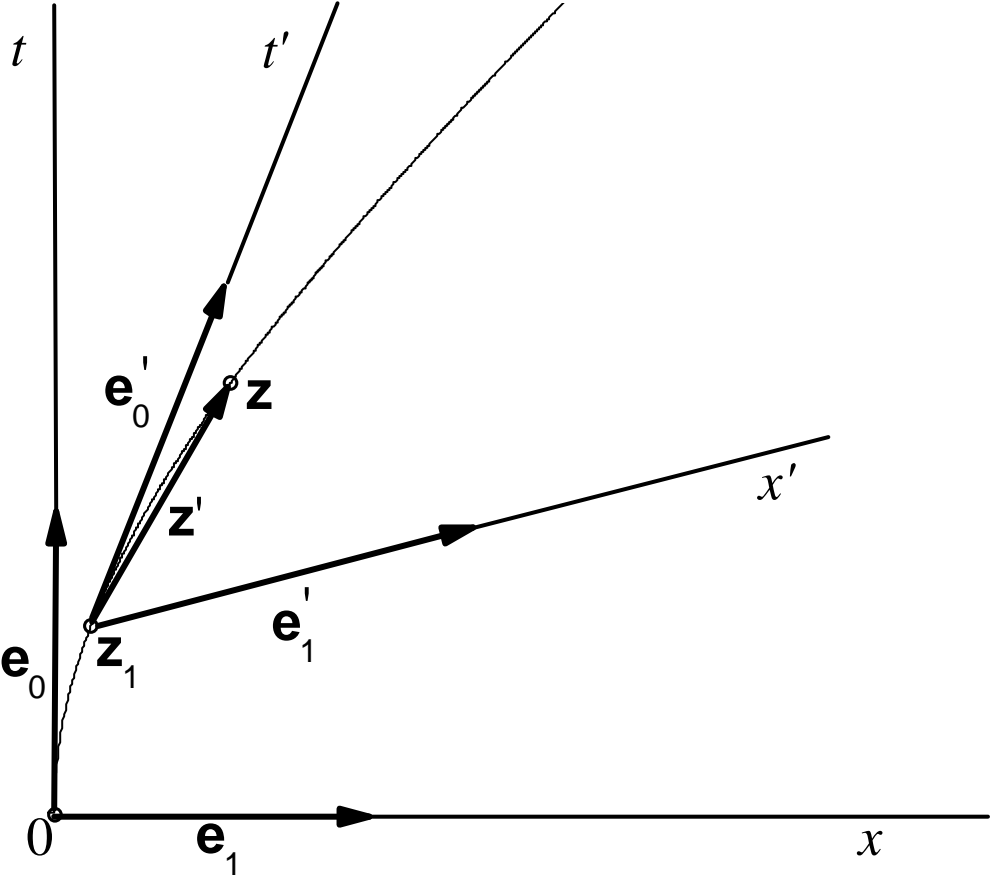